\documentclass[amsmath,amssymb,aps,prr,reprint,longbibliography,superscriptaddress]{revtex4-2}

\usepackage[colorlinks=true,allcolors=blue,breaklinks=true]{hyperref} 
\usepackage{bm} 
\usepackage{bbding}
\usepackage{orcidlink}
\usepackage[normalem]{ulem}

\usepackage{feynmp-auto}
\newcommand{\setval}{
    \fmfset{wiggly_len}{2 mm}
    \fmfset{arrow_len}{3mm}
    \fmfset{arrow_ang}{13}
    \fmfset{dash_len}{2mm}
    \fmfpen{0.15mm}
    \fmfset{dot_size}{1.5thick}
    \fmfcmd{
        style_def Double expr p =
        pickup pencircle scaled 0.25mm; 
        draw_double p;
        pickup pencircle scaled 0.15mm; 
        enddef;
    }
    \fmfcmd{
        style_def response expr p =
        idraw ("plain", subpath (0,length(p)/2) of p);
        idraw ("wiggly", subpath (length(p)/2,length(p)) of p);
        enddef;
    }
    \fmfcmd{%
        vardef cross_bar (expr p, len, ang, place) =
        ((-len/2,0)--(len/2,0))
        rotated (ang + angle direction length(p)/4 of p)
        shifted point length(p)*place of p
        enddef;
    }
    \fmfcmd{
        style_def corr_p expr p =
        cdraw p;
        ccutdraw cross_bar (p, 2mm, 45, 1/2);
        enddef;
    }
    \fmfcmd{
        style_def response_p expr p =
        idraw ("plain", subpath (0,length(p)/2) of p);
        idraw ("wiggly", subpath (length(p)/2,length(p)) of p);
        ccutdraw cross_bar (p, 2mm, 45, 1/2);
        enddef;
    }
    \fmfcmd{%
        style_def corr_p_arrow expr p =
        cdraw p;
        cfill (harrow (p, .75));
        ccutdraw cross_bar (p, 2mm, 45, .25);
        enddef;
    }
    \fmfcmd{%
        style_def corr_p_end expr p =
        cdraw p;
        ccutdraw cross_bar (p, 2mm, 90, 1);
        enddef;
    }
}

\newcommand{\fmfshade}[1]{\fmfv{d.f=shaded,d.shape=circle,d.size=3.5mm}{#1}}

\newcommand{\fmfsquare}[1]{\fmfv{d.f=empty,d.shape=square,d.size=3.5mm}{#1}}

\newcommand{\one}{I}

\usepackage[ISO]{diffcoeff}[=v4]

\newcommand{\fdv}[3][1]{\diff.delta.[#1]{#2}{#3}}
\diffdef{}{long-var-wrap = dv}

\newcommand{\dd}{\mathrm{d}}
\newcommand{\E}[1]{\left \langle #1 \right \rangle}
\newcommand{\Oh}{\mathcal{O}}
\newcommand{\T}{\mathcal{T}}

\newcommand{\D}{\mathcal{D}}
\newcommand{\comma}{,} 
\newcommand{\integral}{I^{(1)}_d}

\newcommand{\Tr}{\mathrm{Tr}\,}

\begin{document}
\begin{fmffile}{feyn/main}

\title{Fluctuation Dissipation Relations for the Non-Reciprocal Cahn-Hilliard Model}

\author{Martin Kj{\o}llesdal Johnsrud\,\orcidlink{0000-0001-8460-7149}
}
\affiliation{Max Planck Institute for Dynamics and Self-Organization (MPI-DS), D-37077 G\"ottingen, Germany}

\author{Ramin Golestanian\,\orcidlink{0000-0002-3149-4002}}
\affiliation{Max Planck Institute for Dynamics and Self-Organization (MPI-DS), D-37077 G\"ottingen, Germany}
\affiliation{Rudolf Peierls Centre for Theoretical Physics, University of Oxford, Oxford OX1 3PU, United Kingdom}

\date{\today}

\begin{abstract}
Recent results demonstrate how deviations from equilibrium fluctuation--dissipation theorem can be quantified for active field theories by deriving exact fluctuations dissipation relations that involve the entropy production [M. K. Johnsrud and R. Golestanian, \href{https://doi.org/10.1103/flzv-lq7x}{Phys. Rev. Res. 7, L032053 (2025)}].
Here we develop and employ diagrammatic tools to perform perturbative calculations for a paradigmatic active field theory, the Non-Reciprocal Cahn-Hilliard (NRCH) model. We obtain analytical results, which serve as an illustration of how to implement the recently developed framework to active field theories, and help to illuminate the specific non-equilibrium characteristics of the NRCH field theory.
\end{abstract}

\maketitle

\textit{Introduction---}Fluctuation dissipation relations (FDRs) are ubiquitous in equilibrium statistical physics.
As consequences of an underlying time-reversal symmetry, they relate the dissipative relaxation of a system to its random fluctuations.
These identities are powerful demonstrations of how fundamental properties at the microscopic level can be leveraged to make statements about the macroscopic behavior of a system. In the same spirit as the relationship between symmetries and conservation laws, invariance of a model under a transformation allows us to derive identities relating correlations and response functions \cite{ZinnJustin2021,aronSymmetriesGeneratingFunctionals2010,siebererThermodynamicEquilibriumSymmetry2015}.
If the system is driven out of equilibrium, these symmetries no longer apply and the restrictions that lead to the FDRs are lifted.
Early examples of generalizing such relations are the fluctuation theorems of Jarzynski~\cite{jarzynskiNonequilibriumEqualityFree1997} and Crooks~\cite{crooksEntropyProductionFluctuation1999}.
Although the symmetry is lost, these relations are still deeply connected to time reversal as they relate the probability of forward- and backward-time paths~\cite{seifertStochasticThermodynamicsFluctuation2012}.
A notable example of a non-equilibrium FDR is the Harada-Sasa relation \cite{haradaEqualityConnectingEnergy2005,haradaEnergyDissipationViolation2006,golestanianPRL2025}, which connects the total deviation from the equilibrium FDR to the entropy production, which is a measure of the irreversibility of the dynamics.

In the emerging subject of active matter~\cite{gompper2020}, an important aim is to connect microscopic dynamics that explicitly break time-reversal symmetry to emergent, macroscopic behavior.
A powerful way to achieve this goal is by using field theories, especially those that can be systematically coarse-grained such that the connection across the scales can be made manifest
~\cite{tonerLongRangeOrderTwoDimensional1995,aditisimhaHydrodynamicFluctuationsInstabilities2002,gelimsonCollectiveDynamicsDividing2015,mahdisoltaniNonequilibriumPolarityinducedChemotaxis2021,BenAlZinati2021,wittkowskiScalarF4Field2014,tjhungClusterPhasesBubbly2018,tiribocchiActiveModelScalar2015,paoluzziScalingEntropyProduction2022,paoluzziNoiseInducedPhaseSeparation2024,sahaScalarActiveMixtures2020,youNonreciprocityGenericRoute2020,pisegnaEmergentPolarOrder2024}.
In these cases, we do not have access to equilibrium FDRs, with a small number of exceptions, such as the Kardar-Parisi-Zhang (KPZ) equation in one dimension~\cite{kardarDynamicScalingGrowing1986}.
A systematic strategy to determine relationships out of thermal equilibrium between fluctuations, response functions, dissipation, and time-reversal is therefore coveted, with examples including efforts such as generalizing the Harada-Sasa relation to active field theories~\cite{nardiniEntropyProductionField2017}.
In a companion paper \cite{johnsrudFluctuationDissipationRelations2025a}, we derive explicit and exact formulas for the deviation from the equilibrium FDR in active field theories by applying time-reversal transformations in the response-field formalism~\cite{tauberCriticalDynamicsField2014,hertzPathIntegralMethods2016b}.
The tools developed there can be used to make powerful statements about active field theories, such as finding a reduced set of equilibrium FDRs that hold for systems with odd-mobility, or shedding light on the suitable definition of entropy production when the symmetries allow for a certain amount of freedom. 

It is important to establish how the formal relations derived in Ref.~\cite{johnsrudFluctuationDissipationRelations2025a} can be implemented in practice. To follow the standard strategy of dealing with field theories \cite{zeeQuantumFieldTheory2003,ZinnJustin2021}, we will naturally need to develop a perturbative diagrammatic framework. By treating simple expectation values in a non-linear model as a series of more involved expectation values in a linear model, one can systematically approximate and evaluate the quantities in question.
Analogously, we can treat time-reversed quantities as a series of time-forward quantities, allowing us to apply the tools of perturbative field theory. The developed framework allows us to derive directly applicable relations between the correlation functions and the response functions in a systematic fashion, yielding generalizations of the equilibrium fluctuation--dissipation theorem (FDT).

In this Letter, we demonstrate how the above strategy can be developed using the specific example of the Non-Reciprocal Cahn-Hilliard (NRCH) model \cite{sahaScalarActiveMixtures2020,youNonreciprocityGenericRoute2020}.
This is a minimal model for interactions that break the action-reaction symmetry~\cite{sotoSelfAssemblyCatalyticallyActive2014,golestanian2024non-reciprocal,uchida2010synchronization,fruchartNonreciprocalPhaseTransitions2021,dinelli2023non,duan2023dynamical,osat2024escaping}, and consequently, time-reversal symmetry, which implies that the FDT does not apply. To apply the formal FDRs found in Ref. \cite{johnsrudFluctuationDissipationRelations2025a} to the NRCH model, we develop a diagrammatic perturbative expansion and obtain relations between correlation and linear response that are valid far from equilibrium. We will first introduce the NRCH model, and then illustrate the techniques in a linear version of the model, before generalizing them to the full non-linear model. While we choose a specific model for the calculations in this paper, the developed approach is directly applicable to other active or otherwise nonequilibrium models \cite{johnsrudFluctuationDissipationRelations2025a}.

\begin{figure}[ht]
    \centering
    \includegraphics[width=\columnwidth]{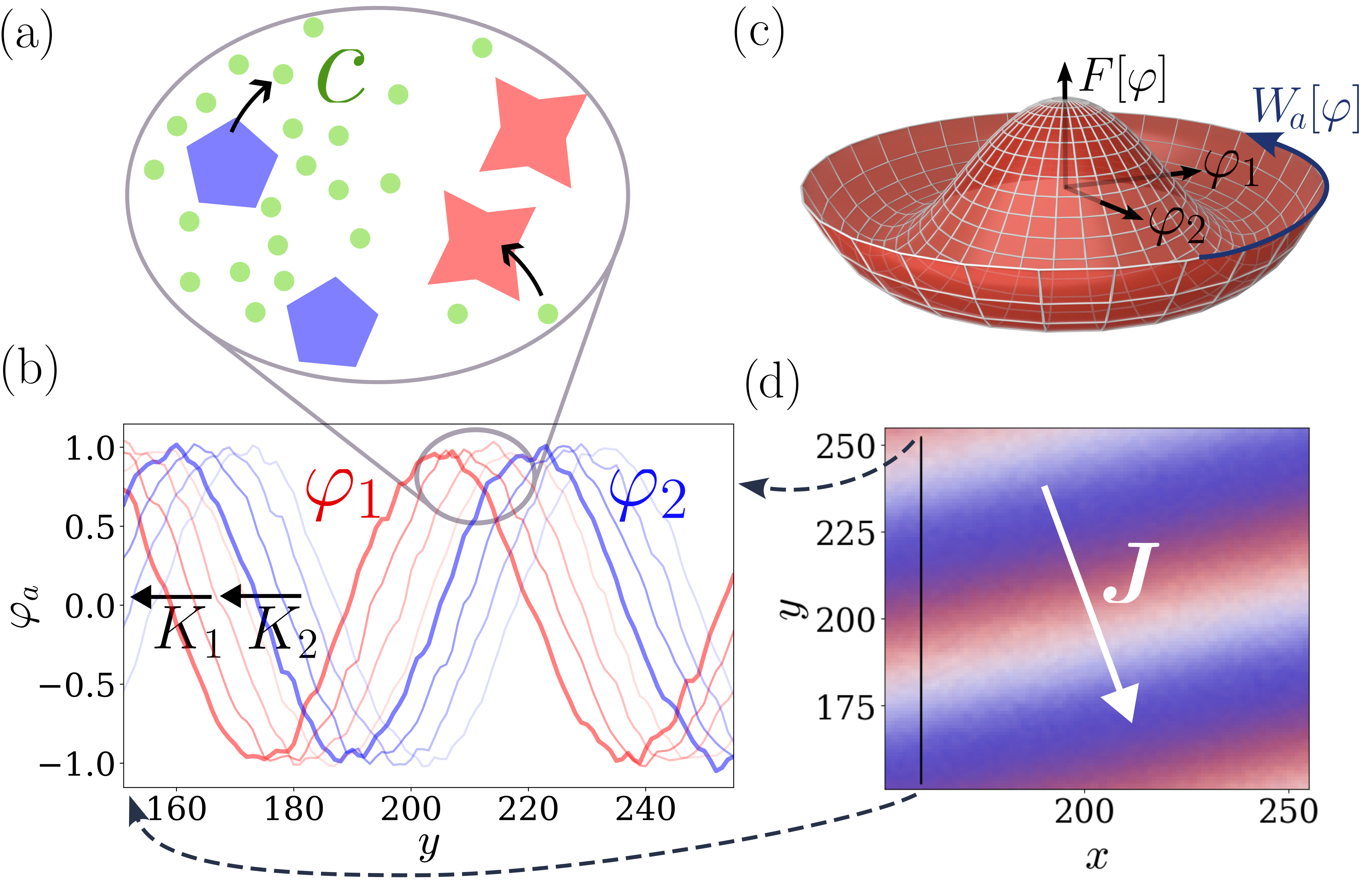}
    \caption{
    Illustration of the NRCH model. (a) At the microscopic level, red and blue particles catalyze chemical reactions that involve the chemical $c$, illustrated in green, which mediates their non-reciprocal interaction \cite{sotoSelfAssemblyCatalyticallyActive2014,agudo-canalejoActivePhaseSeparation2019}. 
    (b) This gives rise to effective interactions $K_a$ acting on the species described by the densities $\varphi_a$. (c) The forces $K_a$ originate from two different sources: the equilibrium component that derives from a free energy $F$, here taken as a Mexican-hat potential [in addition to surface tension term $\propto (\nabla \varphi)^2$], plus non-reciprocal forces $W_a$.
    (d) These forces break time-reversal symmetry, leading to the emergence of a steady state in the form of a traveling pattern with a non-zero current $\bm J$.}
    \label{fig:NRCH}
\end{figure}

\textit{The NRCH model---}%
The effective interactions of living or synthetic agents that constitute nonequilibrium active matter will generically have broken action-reaction symmetry~\cite{sotoSelfAssemblyCatalyticallyActive2014,melzerStructureStabilityPlasma1996,fruchartNonreciprocalPhaseTransitions2021,golestanian2024non-reciprocal}.
As a concrete example, we consider catalytically active particles, which can be propelled by gradients of the reactant and product molecules described by chemical concentration fields $c_i(\bm x, t)$~\cite{golestanian2005propulsion}. The chemical fields then naturally act as mediators of effective interactions between catalytically active particles, because they act as sources and sinks of the chemicals while they simultaneously respond to the chemical gradients $\bm \nabla c_i$ produced by other particles. In mixtures of different types of catalytically active particles, since the rates of the production and consumption of the chemicals and the strengths of the coupling to the gradients generically differ for the different species, their interactions will be non-reciprocal \cite{Golestanian2019phoretic}. In cases where the chemical fields are screened---due to the Michaelis-Menten kinetics of the reaction or decay---the interactions are finite-ranged ~\cite{sahaClustersAstersCollective2014,tucciNonreciprocalCollectiveDynamicsNJP2024}. At the level of dynamical field theory for the density of the catalytically active particle, $\varphi_a(\bm x, t)$, the interactions will manifest themselves in the form of a local effective force, $K_a[\varphi]$, which will generically be non-reciprocal~\cite{sahaScalarActiveMixtures2020}. This is illustrated in Fig.~\ref{fig:NRCH}.

The NRCH model is an extension of the Cahn-Hilliard model~\cite{cahnFreeEnergyNonuniform1958,cahnFreeEnergyNonuniform1959,cahnPhaseSeparationSpinodal1965}, which describes the equilibrium dynamics of phase separation, to systems with broken action-reaction symmetry~\cite{sahaScalarActiveMixtures2020,youNonreciprocityGenericRoute2020}.
It takes the form of a stochastic field equation
\begin{align}
    \label{eq:  model}
    \partial_t \varphi_a(\bm x, t) 
    = - \nabla^2 \Gamma K_a[\varphi](\bm x, t) + \sqrt{- 2\nabla^2 D} \, \xi_a(\bm x, t),
\end{align}
where $\Gamma$ is the mobility and $D$ is the noise-strength. The force
can be represented as follows
\begin{align}
    \label{eq:  force}
    K_a[\varphi]
    = - \fdv{F[\varphi]}{{\varphi_a}}+W_a[\varphi],
\end{align}
where $F$ is the free energy, and $W_a[\varphi]$ is a non-reciprocal interaction. For two species, the latter takes the form 
\begin{align}
    \label{eq:  NR force}
W_a[\varphi] = g[\varphi] \epsilon_{ab}\varphi_b, 
\end{align}
where $\epsilon$ is the anti-symmetric Levi-Civita tensor obeying $\epsilon_{12} = -\epsilon_{21} = 1$, and $g[\varphi]$ is a functional of $\varphi_a$. Summation over repeated indices is implied. This form cannot be written as a functional derivative of a free energy, thus breaking time-reversal symmetry. Equations \eqref{eq:  force} and \eqref{eq:  NR force} represent the Helmholtz-decomposition of the force in $\varphi_a$-space into its divergence-free and curl-free parts, as $\int_{\bm x} \fdv{W_a}{\varphi_a} = 0$ and $\int_{\bm x} \epsilon_{ab} \fdv{}{\varphi_a} \fdv{F}{\varphi_b} = 0$. Moreover, Eqs.~\eqref{eq:  force} and \eqref{eq:  NR force} give rise to a steady-state entropy production of the form 
\begin{align}
    \label{eq: entropy}
    S[\varphi] = 
    \frac{\Gamma}{D} 
    \int_{t,\bm x} \partial_t \varphi_a(\bm x, t) 
    W_a[\varphi](\bm x, t),
\end{align}
where we have used the shorthand $\int_{t, \bm x} \equiv \int \dd t \int \dd^d {\bm x}$.

The NRCH features traveling waves~\cite{sahaScalarActiveMixtures2020,youNonreciprocityGenericRoute2020} that represent an underlying true long-range polar order even in 2D \cite{pisegnaEmergentPolarOrder2024}, effervescent phases~\cite{sahaEffervescentWavesBinary2022}, topological defects~\cite{rana2024a,rana2024b}, as well as complex spatiotemporal patterns in multi-component systems \cite{parkavousiEnhancedStabilityChaotic2025}. Entropy production in NRCH has also been the subject of recent investigations \cite{LoosNRCH2023,LucaNRCH2023}.
The model is taken to be invariant under rotations in $\varphi_a$ (i.e. $\mathrm{SO}(2)$ symmetry)~\cite{sahaEffervescentWavesBinary2022,rana2024a, pisegnaEmergentPolarOrder2024}, as illustrated in Fig.~\ref{fig:NRCH}, since both $F$ and $g$ depend only on $\varphi^2 = \varphi_1^2 + \varphi_2^2$. Invoking this symmetry renders NRCH theory as the conserved generalization of the Complex Ginzburg-Landau equation with noise~\cite{aransonWorldComplexGinzburgLandau2002}.
This restricted model serves to simplify analysis, while it has still been shown to feature the main non-equilibrium phenomenology of the NRCH~\cite{pisegnaEmergentPolarOrder2024,rana2024a,sahaEffervescentWavesBinary2022}.

An important identity for systems with an equilibrium steady state, such as the Cahn-Hilliard model, is the FDT. It relates correlation functions, defined in real space as $C_{ab}(\bm x-\bm x',t-t')\equiv\E{\varphi_a(\bm x, t) \varphi_b(\bm x', t')}$, and the linear response of the system to a perturbation field $h_b(\bm x, t)$, quantified by the susceptibility $\chi_{ab}$.
This is defined in Fourier space as $\chi_{ab}(\bm q, \omega) \delta_{\bm q+\bm q'} \delta_{\omega +\omega'} \equiv \fdv{\E{\varphi_a(\bm q, \omega)}}{{h_b(\bm q',\omega')}}$
\footnote{We denote $\delta_{\bm q+\bm q'} \equiv (2 \pi)^{d} \delta^d(\bm q + \bm q') $ and $\delta_{\omega + \omega'}\equiv 2 \pi  \delta(\omega + \omega')$ for brevity.}.
The susceptibility is related to the Green's function $G$ via $\chi = q^2 \Gamma G$.
Interestingly, the correlation functions and the Green's functions are related as follows 
\begin{align}
    \label{eq: CGG causality}
C_{ab}(\bm q, \omega) = 2q^2D [G G^\dagger]_{ab}(\bm q, \omega),
\end{align}
even in non-linear nonequilibrium systems, as a consequence of causality \cite{supp,tauberCriticalDynamicsField2014,hertzPathIntegralMethods2016b}.
The power of the FDT is that $C$ and $\chi$ (or $G$) are important observables of the system, both experimentally and theoretically. Lack of such stringent criteria is one of the reasons generalizing statistical mechanics to systems far from equilibrium is such a daunting task. Here, by applying methods from perturbative field theory on the FDRs proposed in Ref.~\cite{johnsrudFluctuationDissipationRelations2025a}, we derive the following identity
\begin{align}
   &
    G_{ab}(\bm q, \omega) 
    -
    C_{ac}(\bm q, \omega) 
    {C_{cd}^{-1}}(-\bm q, -\omega)\,
    G_{db}(- \bm q, - \omega)
    \nonumber\\
    &\hskip1cm
     = 
    \frac{i\omega}{q^2 \! D} C_{ab}(\bm q, \omega), \label{eq: FDR}
\end{align}
using a one-loop approximation. This relation, which is the main result of the paper, is as stringent as the FDT; in fact, an FDT that is valid even in non-equilibrium conditions is recovered in the reciprocal limit where $C = C^*$. The result provides a powerful identity for the NRCH class of active matter, which can be used in experimental or computational probes of the behavior of such systems.

As quantitative probes, $C$ and $G$ can either be considered phenomenological at the field-theory level and measured directly, or connected to a microscopic model. In the former case, the correlation function is measured from a time-series of the densities $\varphi_a$, while $G$ (or the response to perturbations $\chi$) can either be measured by perturbing the system or indeed inferred from Eq.~\eqref{eq: FDR} given $C$. The relation holds for any and all frequencies or wave-numbers. In the latter case, these quantities are functions of the microscopic parameters accessible to a coarse-graining procedure. For example, the system of catalytically active particles illustrated in Fig.~\ref{fig:NRCH} can be coarse-grained to tie the microscopic dynamics to the parameters in the field theory~\cite{tucciNonreciprocalCollectiveDynamicsNJP2024}.
Moreover, the derivation of Eq.~\eqref{eq: FDR} demonstrates how our framework can be used in practice to derive identities in other non-equilibrium systems by applying diagrammatic techniques. 

\textit{FDRs---}%
The action associated with Eq.~\eqref{eq:  model} is~\cite{tauberCriticalDynamicsField2014}
\begin{align}\label{eq:  action}
    &A[\varphi, \tilde \varphi]  = 
    \!\int\limits_{t, \bm x}\!
    \left\{
        i \tilde \varphi_a
        \Big(
            \partial_t \varphi_a + \nabla^2 \Gamma K_a
        \Big)
        - \tilde \varphi_a \nabla^2 D \tilde \varphi_a
        \right\},
\end{align}
where $\tilde \varphi_a$ is a response-field (or auxiliary field) used to incorporate the noise averaging at the outset. We define a vector of the response and physical fields, $(\psi_1,\cdots) = (i\tilde\varphi_1, \cdots ,\varphi_1,\cdots)$.
In general, we use the indices $a, b, c\dots$ when referring to $\varphi$ and $\tilde \varphi$, while $i,j,k\dots$ refer to the combination $\psi$.
Matrices acting on $\psi$ are block-matrices acting on $\tilde \varphi_a$ and $\varphi_a$.
We use the shorthand notation
\begin{align}
    M_{ij}\psi_j
    = 
    \begin{pmatrix}
        M_{\tilde a \tilde b} & M_{\tilde a b}\\
        M_{a \tilde b} & M_{a b}
    \end{pmatrix}
    \begin{pmatrix}
        i \tilde \varphi_b \\
        \varphi_b
    \end{pmatrix},
\end{align}
and refer to the sub-matrices as ``quadrants''. The tilde on the subscript (such as $M_{a \tilde b}$) indicate that this element acts on $i\tilde\varphi_{b}$.
The expectation value of a functional $\Oh[\psi]$ is 
$
    \E{\Oh[\psi]} = \int \D \psi \, \Oh[\psi] e^{-A[\psi]}
$.

In Ref. \cite{johnsrudFluctuationDissipationRelations2025a}, we obtain four exact matrix identities:
\begin{align}
    \label{eq: GFDT matrix} 
    \hspace{-1.5mm}
    \begin{bmatrix}
        \frac{i \omega}{D q^2}\!\!
        \left(
            G_+
            \!\!- G_+^\dagger
            \!- \!\frac{i \omega}{D q^2} C_+\!
        \right)
        &
        G_+^\dagger
        \!\!- G_-^\dagger
        \!+ \!\frac{i \omega}{D q^2} C_+^\dagger
        \\[.22cm]
        G_+
        \!\!- G_-
        \!- \!\frac{i \omega}{D q^2} C_+ 
        & C_+ \!- C_-
    \end{bmatrix}\!
    =\!
    \Delta_-.\!\!\!
\end{align}
Here, the sign in the subscript denotes the sign of the arguments, e.g., $C_{\pm, ab}(\bm q, \omega) = C_{ab}(\pm \bm q, \pm \omega)$.
The Hermitian conjugation is defined as $G_\pm^\dagger \equiv \left(G_\pm\right)^\dagger = G_\mp^T$, and we note that in Fourier space we have $G_- = G_+^*$.
The right-hand side of Eq. \eqref{eq: GFDT matrix} is 
\begin{align}
  \hskip-.1cm
   \E{\psi_i(\bm q, \omega) \psi_j(\bm q'\!, \omega')\left(e^{\!-S}\! - 1\right)}  
   \!\equiv\! 
   \Delta_{ij}(\bm q, \omega)   \delta_{\bm q+\bm q'} \delta_{\omega +\omega'}\!,
  \hskip-.1cm
\end{align}
where $S$ is the total entropy production operator, defined in Eq. \eqref{eq: entropy} for NRCH.

\textit{Linear model}---To introduce the diagrammatic approach and demonstrate how it works in practice, we start by including only the linear terms in the equation of motion. In this case, the interaction term reads
$
    K_a[\varphi]
    = 
    -\left[
    \left( r - \nabla^2\right) \delta_{ab} + \alpha_0 \epsilon_{ab}
    \right] \varphi_b
$, corresponding to $F = \frac{1}{2} \int_{\bm x} \varphi_a(r - \nabla^2)\varphi_a$, and $g = - \alpha_0$.
We consider the FDR given in the lower quadrants of Eq.~\eqref{eq: GFDT matrix}. For the right-hand side of Eq.~\eqref{eq: GFDT matrix}, we employ a diagrammatic approach that we develop as follows. The propagators are drawn as
$C_{ab} =
\parbox{14mm}{
\centering
\begin{fmfgraph*}(10,3)
    \setval
    \fmfleft{i}
    \fmfright{o}
    \fmf{plain}{i,o}
    \fmfv{l=$a$,l.d=.05w}{i}
    \fmfv{l=$b$,l.d=.05w}{o}
\end{fmfgraph*}
}
$
and
$G_{ab} =
\parbox{14mm}{
\centering
\begin{fmfgraph*}(10,3)
    \setval
    \fmfleft{i}
    \fmfright{o}
    \fmf{plain}{i,c}
    \fmf{wiggly}{c,o}
    \fmfv{l=$a$,l.d=.05w}{i}
    \fmfv{l=$b$,l.d=.05w}{o}
\end{fmfgraph*}
}
$ \cite{tauberCriticalDynamicsField2014}. 
The response propagator is obtained by inverting the equations of motion, which in this case can be done explicitly, yielding
\begin{align}
    G = -i \omega\one + q^2 \Gamma \left[(r + q^2) \one + \alpha_0 \epsilon\right], \label{eq: G}
\end{align}
where $I$ denotes the identity tensor. The correlation propagator can be derived using $C = 2q^2DGG^\dagger$.

We can, at this point, calculate $C$ explicitly in terms of the parameters $r$ and $\alpha_0$ to verify Eq.~\eqref{eq: FDR}, as is done in \cite{supp}; see Eq.~(S17) and the discussion leading up to it.
We may furthermore directly tie it to microscopic models such as the catalytic active particles, illustrated in Fig.~\ref{fig:NRCH}.
As an example, in Ref.~\cite{tucciNonreciprocalCollectiveDynamicsNJP2024} $\alpha_0$ ($\psi_0$ in their notation) is expressed explicitly [see their Eq.~(31)] in terms of microscopic parameters such as the density of the two species, catalytic activity, mobility and so on.
These can further be related to specific experimental setups, such as Janus-colloids catalyzing hydrogen peroxide; see Table 1 of Ref.~\cite{tucciNonreciprocalCollectiveDynamicsNJP2024}.

We now introduce the tools to derive Eq.~\eqref{eq: FDR}.
By writing the total entropy production [Eq.~\eqref{eq: entropy}] in the following form $S = - \int_{\bm q, \omega} \frac{1}{2} \varphi_a \sigma_{ab}\varphi_b^*$, we next introduce a vertex corresponding to ``entropy consumption'', namely,
\begin{align}
    \parbox{16mm}{
    \centering
    \begin{fmfgraph*}(12,5)
        \setval
        \fmfleft{i}
        \fmfright{o}
        \fmf{plain}{i,c}
        \fmf{plain}{c,o}
        \fmfv{l=$a$,l.d=.05w}{i}
        \fmfv{l=$b$,l.d=.05w}{o}
        \fmfdot{c}
        \fmffreeze
        \fmf{phantom}{i,d1,c}
        \fmfv{l=$\omega$,l.d=.1w,l.a=90}{d1}
        \fmf{phantom}{c,d2,o}
        \fmfv{l=$\omega'$,l.d=.05w,l.a=-90}{d2}
    \end{fmfgraph*}
    }
    \equiv
    \sigma_{ab}(\omega)\delta_{\omega+\omega'}
    = 
    - 2 i \omega  \Gamma \frac{ \alpha_0}{ D } \epsilon_{ab}\delta_{\omega+\omega'}.
\end{align}
Note that the form of this vertex is model-specific, as can be seen e.g. from the fact that here it only has two legs, since the active term is linear.

The lower-left quadrant of the right-hand side of Eq.~\eqref{eq: GFDT matrix} is denoted $\Delta_{-,a\tilde b}$.
Expanding this as a perturbative series in the non-reciprocal parameter $\alpha_0$ (using the diagrammatic rules introduced above) yields
\begin{widetext}
\begin{align}
    \label{eq: linear expansion}
    \Delta_{a \tilde b}
    &=
    \parbox{10mm}{
    \begin{fmfgraph*}(10,5)
        \setval
        \fmfleft{i}
        \fmfright{o}
        \fmf{plain}{i,c}
        \fmf{plain,tension=2}{c,k}
        \fmf{wiggly,tension=2}{k,o}
        \fmfdot{c}
    \end{fmfgraph*}
    }
    +
    \parbox{8mm}{
    \centering
    \begin{fmfgraph*}(5,0)
        \setval
        \fmfleft{i}
        \fmfright{o}
        \fmf{plain,left}{i,o}
        \fmf{plain,left}{o,i}
        \fmfdot{o}
    \end{fmfgraph*}
    \begin{fmfgraph*}(8,1)
        \setval
        \fmfleft{i}
        \fmfright{o}
        \fmf{plain}{i,k}
        \fmf{photon}{k,o}
    \end{fmfgraph*}
    }
    + 
    \frac{1}{2!}
    \bigg(
    \parbox{12mm}{
    \begin{fmfgraph*}(12,5)
        \setval
        \fmfleft{i}
        \fmfright{o}
        \fmf{plain}{i,c1}
        \fmf{plain}{c1,c2}
        \fmf{plain,tension=2}{c2,k}
        \fmf{photon,tension=2}{k,o}
        \fmfdot{c1}
        \fmfdot{c2}
    \end{fmfgraph*}
    }
    +
    \parbox{10mm}{
    \centering
    2
    \begin{fmfgraph*}(4,2)
        \setval
        \fmfleft{i}
        \fmfright{o}
        \fmf{plain,left}{i,o}
        \fmf{plain,left}{o,i}
        \fmfdot{o}
    \end{fmfgraph*}
    \\
    \begin{fmfgraph*}(10,-3)
        \setval
        \fmfleft{i}
        \fmfright{o}
        \fmf{plain}{i,c}
        \fmf{plain,tension=2}{c,k}
        \fmf{wiggly,tension=2}{k,o}
        \fmfdot{c}
    \end{fmfgraph*}}
    +
    \parbox{17mm}{
    \centering
    \begin{fmfgraph*}(3.5,2)
        \setval
        \fmfleft{i}
        \fmfright{o}
        \fmf{plain,left}{i,o}
        \fmf{plain,left}{o,i}
        \fmfdot{o}
        \fmfdot{i}
    \end{fmfgraph*}
    +\hspace{-.01cm}2\hspace{-.06cm}
    ${
    \begin{fmfgraph*}(3.5,2)
        \setval
        \fmfleft{i}
        \fmfright{o}
        \fmf{plain,left}{i,o}
        \fmf{plain,left}{o,i}
        \fmfdot{o}
    \end{fmfgraph*}}^{\,2}$
    \\
    \begin{fmfgraph*}(12,-4)
        \setval
        \fmfleft{i}
        \fmfright{o}
        \fmf{plain}{i,k}
        \fmf{photon}{k,o}
    \end{fmfgraph*}
    }
    \bigg)
    +
    \frac{1}{3!}
    \bigg(
    \parbox{15mm}{
    \begin{fmfgraph*}(15,5)
        \setval
        \fmfleft{i}
        \fmfright{o}
        \fmf{plain}{i,c1}
        \fmf{plain}{c1,c2}
        \fmf{plain}{c2,c3}
        \fmf{plain,tension=2}{c3,k}
        \fmf{photon,tension=2}{k,o}
        \fmfdot{c1}
        \fmfdot{c2}
        \fmfdot{c3}
    \end{fmfgraph*}
    }
    +
    \parbox{12mm}{
    \centering
    3
    \begin{fmfgraph*}(4,2)
        \setval
        \fmfleft{i}
        \fmfright{o}
        \fmf{plain,left}{i,o}
        \fmf{plain,left}{o,i}
        \fmfdot{o}
    \end{fmfgraph*}
    \begin{fmfgraph*}(12,-3)
        \setval
        \fmfleft{i}
        \fmfright{o}
        \fmf{plain}{i,c1}
        \fmf{plain}{c1,c2}
        \fmf{plain,tension=2}{c2,k}
        \fmf{photon,tension=2}{k,o}
        \fmfdot{c1}
        \fmfdot{c2}
    \end{fmfgraph*}
    }
    +\cdots
    \bigg)
    + \cdots, 
\end{align}\label{eq: diagrams-1}
which simplifies to 
\begin{align}
\Delta_{a \tilde b}= 
    -\,
    \parbox{8mm}{
    \centering
    \begin{fmfgraph*}(8,5)
        \setval
        \fmfleft{i}
        \fmfright{o}
        \fmf{plain}{i,k}
        \fmf{photon}{k,o}
    \end{fmfgraph*}
    }
    +
    \Big(
    \parbox{8mm}{
    \centering
    \begin{fmfgraph*}(8,5)
        \setval
        \fmfleft{i}
        \fmfright{o}
        \fmf{plain}{i,k}
        \fmf{photon}{k,o}
    \end{fmfgraph*}
    } 
    +
    \parbox{10mm}{
    \begin{fmfgraph*}(10,5)
        \setval
        \fmfleft{i}
        \fmfright{o}
        \fmf{plain}{i,c}
        \fmf{plain,tension=2}{c,k}
        \fmf{wiggly,tension=2}{k,o}
        \fmfdot{c}
    \end{fmfgraph*}
    }
    +
    \parbox{12mm}{
    \begin{fmfgraph*}(12,5)
        \setval
        \fmfleft{i}
        \fmfright{o}
        \fmf{plain}{i,c1}
        \fmf{plain}{c1,c2}
        \fmf{plain,tension=2}{c2,k}
        \fmf{photon,tension=2}{k,o}
        \fmfdot{c1}
        \fmfdot{c2}
    \end{fmfgraph*}
    }
    +
    \parbox{15mm}{
    \begin{fmfgraph*}(15,5)
        \setval
        \fmfleft{i}
        \fmfright{o}
        \fmf{plain}{i,c1}
        \fmf{plain}{c1,c2}
        \fmf{plain}{c2,c3}
        \fmf{plain,tension=2}{c3,k}
        \fmf{photon,tension=2}{k,o}
        \fmfdot{c1}
        \fmfdot{c2}
        \fmfdot{c3}
    \end{fmfgraph*}
    }
    + \cdots
    \Big)
    \Big(
    1 + 
    \parbox{5mm}{
    \begin{fmfgraph*}(5,5)
        \setval
        \fmfleft{i}
        \fmfright{o}
        \fmf{plain,left}{i,o}
        \fmf{plain,left}{o,i}
        \fmfdot{o}
    \end{fmfgraph*}
    }
    +
    \frac{1}{2!}
    \Big[ 
    \parbox{5mm}{
    \begin{fmfgraph*}(4,4)
        \setval
        \fmfleft{i}
        \fmfright{o}
        \fmf{plain,left}{i,o}
        \fmf{plain,left}{o,i}
        \fmfdot{o}
        \fmfdot{i}
    \end{fmfgraph*}
    }
    +2\,
    \parbox{5mm}{
    ${
    \begin{fmfgraph*}(4,4)
        \setval
        \fmfleft{i}
        \fmfright{o}
        \fmf{plain,left}{i,o}
        \fmf{plain,left}{o,i}
        \fmfdot{o}
    \end{fmfgraph*}}^2
    $}\Big]
    +\cdots
    \Big),\label{eq: diagrams}
   \end{align}
\end{widetext}
because the vacuum bubbles---diagrams without external legs---factorize. 
These bubbles sum up to $\E{e^{-S}} = 1$ (see \cite{supp} for details). 
The remaining connected diagrams are easily calculated, as we illustrate by writing out the simplest diagram
\begin{align}
    &
    \fmfframe(3,0)(2,-1.1){
    \begin{fmfgraph*}(18,5)
        \setval
        \fmfleft{i}
        \fmfright{o}
        \fmf{plain}{i,l}
        \fmf{plain}{l,c}
        \fmf{plain}{c,k}
        \fmf{photon}{k,o}
        \fmfdot{c}
        \fmfv{l=$a$,l.d=.05w}{i}
        \fmfv{l=$b$,l.d=.05w}{o}
        \fmffreeze
        \fmf{phantom}{i,d1,c}
        \fmfv{l=${\bm{q'}\comma\omega'}$,l.d=.05w,l.a=90}{d1}
        \fmf{phantom}{c,d2,o}
        \fmfv{l=${\bm{q}\comma\omega}$,l.d=.05w,l.a=-90}{d2}
    \end{fmfgraph*}
    } 
    \nonumber\\
        &=
    \int\limits_{\nu, \bm k}
    C_{ac}(\bm q', \omega')
    \delta_{\omega' + \nu}
    \delta_{\bm q' + \bm k}
    \sigma_{cd}(\nu)
    G_{db}(-\bm k, -\nu)
    \delta_{\nu-\omega}
    \delta_{\bm k-\bm q},
    \nonumber \\
    & = 
    C_{ac}(- \bm q, - \omega) \sigma_{cd}(\omega)G_{db}(- \bm q, - \omega) \delta_{\bm q + \bm q'}\delta_{\omega + \omega'}.
\end{align}
We observe that diagrams with two external legs and no loops result in the chaining together of propagator and entropy-consumption vertices, with the signs of the arguments alternating, yielding
\begin{align}
    \Delta_-
    & = C_- \sigma_+ G_- + C_-\sigma_+ C_- \sigma_+ G_- \nonumber \\
    & \quad\quad + C_-\sigma_+ C_- \sigma_+ C_- \sigma_+ G_-
    + \cdots, \nonumber \\
    & =  C_-\sigma_+ \left[ \one + C_- \sigma_+  + C_- \sigma_+ C_- \sigma_+  + \cdots \right] G_-.
\end{align}
The result is summed as a Dyson series, resulting in $\Delta_{-,a\tilde b} = [C_- \Sigma_+ G_-]_{a b}$, where we defined a re-summed entropy operator
\begin{align}\label{eq: ressuemd sigma}
    \Sigma(\bm q, \omega) 
    \equiv \left[\sigma(\omega)^{-1} - C(- \bm q, -\omega)\right]^{-1}.
\end{align}
The same techniques can be applied to the other blocks of Eq.~\eqref{eq: GFDT matrix}.
$\Sigma$ corresponds to the amputated version of $\Delta$, and will therefore appear also there, e.g. for the antisymmetric part of the correlation function,
$
    C_+ - C_-
    = C_- \Sigma_+ C_-
$
(note that $C^\dagger = C$, so $C_- = C^T_+$). A nonvanishing antisymmetric part of $C$ and the deviation from the FDT are both signatures of nonequilibrium behavior, and our formalism connects them through $\Sigma$.
All these relations can be verified by explicit calculation of $C$ and $G$, as detailed in~\cite{supp}.

We have further verified our scheme by explicitly calculating the right-hand side of Eq.~\eqref{eq: GFDT matrix} in terms of a new action $A_S = A + S$, for which $\D_{ij}' \equiv \E{\psi_i \psi_j e^{-S}}$ is the propagator, and subsequently using $\Delta  = \D' - \D$, where $\D_{ij}\equiv \E{\psi_i \psi_j}$.
This method, however, becomes significantly more cumbersome for more complex, nonlinear models, as some of the nice features of the response-field formalism are lost. 
In particular, the response-response block $\D_{R,\tilde a \tilde b}' = \E{\tilde \varphi_a \tilde \varphi_be^{-S}}$ will not vanish (unlike $\D_{\tilde a \tilde b}$).

\textit{Nonlinear model}---The diagrammatic approach allows us to perform calculations in richer theories.
We extend the NRCH by adding a quartic term, $\frac{1}{4} u \varphi^4$, to the free energy. We also add a nonlinear term, $-\alpha_1 \varphi^2 \epsilon_{ab} \varphi_b$, and a surface-tension, $\beta \nabla^2 \epsilon_{ab}\varphi_b$, to the nonreciprocal interaction $W_a$.
The total interaction is then
\begin{align}
    \hspace{-.08cm}
    K_a\! =\! - (r\! - \!\nabla^2\! + \!u \varphi^2 )\varphi_a\! - \!(\alpha_0\! - \! \beta_0 \nabla^2\! + \!\alpha_1 \varphi^2)\epsilon_{ab}\varphi_b.\!\!
\end{align}
This version of NRCH contains all the relevant terms (in the Renormalization Group sense) with rotational invariance~\cite{johnsrudPhaseDiagramNonReciprocal2025,johnsrudRenromalizationTBP}, and has been used as a testing ground for exploring the effects of nonreciprocity in conserved systems~\cite{sahaEffervescentWavesBinary2022,rana2024a,rana2024b}. 
The free energy and traveling wave state are illustrated Fig. \ref{fig:NRCH}.

The Green's function for the linearized dynamics is now given by $G = -i \omega\one + q^2 \Gamma \left[(r + q^2) \one + (\alpha_0 + \beta_0 q^2) \epsilon\right]$. Moreover, the response-action $A[\psi]$ acquires the additional interaction term
\begin{align}
    A_\mathrm{I}[\psi] 
    = - \int_{t, \bm x} q^2\Gamma g_{abcd} \;i \tilde \varphi_a\varphi_b\varphi_c\varphi_d,
\end{align}
where $g_{abcd} = (u \delta_{ab} + \alpha_1 \epsilon_{ab})\delta_{cd}$.
The entropy is now
\begin{align}
    S 
    = 
    -
    \int_{t, \bm x}
    \left(
        \frac{1}{2}
        \sigma^{(2)}_{ab} \varphi_a \varphi_b
        + 
        \sigma^{(4)}_{abcd}\varphi_a \varphi_b \varphi_c\varphi_d
    \right).
\end{align}
where we have defined the entropy consumption couplings
\begin{align}
    \!\sigma^{(2)}_{ab}\! = \!- 2 i\omega\Gamma \frac{\alpha_0 + \beta_0 q^2}{D}   \epsilon_{ab},\,\,
    \sigma^{(4)}_{abcd}\! = \!-i \omega\Gamma \frac{\alpha_1}{D}  \epsilon_{ab} \delta_{cd},
\end{align}
representing a four-point vertex, in addition to the original two-point vertex.
These new terms are represented graphically by four-point vertices as follows
\begin{align}
    \hspace{-.2cm}
    \parbox{16mm}{
    \fmfframe(4,1)(0,1){ 
    \begin{fmfgraph*}(14,12)
        \setval
        \fmfleft{i1,i2}
        \fmfright{o1,o2}
        \fmfv{l=$a$,l.d=.1w}{i1}
        \fmfv{l=$b$,l.d=.1w}{i2}
        \fmfv{l=$c$,l.d=.1w}{o1}
        \fmfv{l=$d$,l.d=.1w}{o2}
        \fmf{wiggly}{i1,v}
        \fmf{plain}{v,o1}
        \fmf{plain}{i2,v,o2}
        \fmfv{l=$\bm q$,l.a=-112,l.d=.25w}{v}
    \end{fmfgraph*}
    }}
    & = -q^2 \Gamma g_{a(bcd)}^{(4)},
    &
    \hspace{-.5cm}
    \parbox{16mm}{
    \fmfframe(4,1)(0,1){
    \begin{fmfgraph*}(14,12)
        \setval
        \fmfleft{i1,i2}
        \fmfright{o1,o2}
        \fmfv{l=$a$,l.d=.1w}{i1}
        \fmfv{l=$b$,l.d=.1w}{i2}
        \fmfv{l=$c$,l.d=.1w}{o1}
        \fmfv{l=$d$,l.d=.1w}{o2}
        \fmf{fermion}{i1,v}
        \fmf{plain}{v,o1}
        \fmf{plain}{i2,v,o2}
        \fmfdot{v}
        \fmfv{l=$\omega$,l.a=-112,l.d=.3w}{v}
    \end{fmfgraph*}
    }}
    & = \sigma^{(4)}_{a(bcd)}.
\end{align}
The first vertex is the standard interaction-vertex in dynamical field theory, while the second is a new addition, which represents the interaction needed for the time-reversed expectations.
The arrow symbolizes the leg of the frequency $\omega$, and the parenthesis around the indices indicate symmetrization.
(We choose not to symmetrize with respect to $a$ to avoid having a complicated $\omega$-dependence.)

Following standard treatments of perturbative field theory, we systematically approximate quantities, such as the ``renormalized'' Green's function $G_R$ for the full nonlinear equations in a loop-expansion.

At the one-loop level, we may write
\begin{align}
    &\Delta_{ij}
    =
    \E{
        \varphi_i \varphi_j
        \left(e^{-S_2} - 1\right)
    }
    \overset{\text{1-loop}}{\sim}
    \Delta_{ij}^{(1)}
    + \Delta_{ij}^{(2)}
    \\\nonumber
    &\equiv
    \E{
        \psi_i\psi_j
        \left(e^{-S_2} - 1\right)e^{-A_\mathrm{I}}
    }_0
    +
    \E{
    \psi_i\psi_j
    \left(e^{-S_4} - 1\right)e^{-S_2}
    }_0,
\end{align}
where expectation values with a zero subscript are taken in the linear theory.
We have here defined $S = S_2 + S_4$, where $S_2$ is the entropy consumption associated with $\sigma^{(2)}$, and $S_4$ derives from $\sigma^{(4)}$.
The terms $\Delta^{(1)}$ and $\Delta^{(2)}$ include only one of the four-point vertices each, which simplifies the calculations.

We will first consider $\Delta^{(1)}$, which means we only have the two-point entropy-consumption vertex, but we have the standard four-point interaction vertex representing $g^{(4)}$. The corrections to $G$ and $C$ are given by the self energy, for which the leading order (one-loop) correction is 
$    
\parbox{10mm}{
\centering
\begin{fmfgraph*}(10,3.3)
    \setval
    \fmfforce{0w,0.1h}{i}
    \fmfforce{1w,0.1h}{o}
    \fmftop{t}
    \fmf{wiggly}{i,c}
    \fmf{plain}{o,c}
    \fmffreeze
    \fmf{plain,left}{c,t}
    \fmf{plain,left}{t,c}
\end{fmfgraph*}
}
$.
Through a Dyson series structure, this gives rise to the renormalized propagators 
$
\E{\varphi \varphi} = 
\parbox{10mm}{
\centering
\begin{fmfgraph*}(10,4)
    \setval
    \fmfleft{i}
    \fmfright{o}
    \fmf{plain}{i,v,o}
    \fmfv{d.f=hatched,d.shape=circle,d.size=3.5mm}{v}
\end{fmfgraph*}
}
$
and
$
\E{\varphi i \tilde \varphi} = 
\parbox{10mm}{
\centering
\begin{fmfgraph*}(10,4)
    \setval
    \fmfleft{i}
    \fmfright{o}
    \fmf{plain}{i,v}
    \fmf{wiggly}{v,o}
    \fmfv{d.f=hatched,d.shape=circle,d.size=3.5mm}{v}
\end{fmfgraph*}
}
$, with renormalized $r$ and $\alpha_0$-parameters, namely, $r_R = r + \delta r$ and $\alpha_{0,R} = \alpha_0 + \delta \alpha_0$, which are the only parameters that are renormalized to one-loop order. These calculations are detailed in~\cite{supp}.
As we only consider the FDR between two-point functions here, we need not consider the renoramlization of $u$ and $\alpha_1$ \cite{johnsrudRenromalizationTBP}.

The diagrammatic expansion for $\Delta$ is significantly more complicated when we include interactions. 
We find that a large portion of the diagrams corresponds to the renormalization of the propagators $C$ and $G$, and all corrections to $\Delta^{(1)}$ above and beyond this are contained in a renormalized entropy-consumption vertex 
$
\sigma_{R,ij}^{(2)}
\equiv
\parbox{10mm}{
\centering
\begin{fmfgraph*}(10,5)
    \setval
    \fmfleft{i}
    \fmfright{o}
    \fmf{Double}{i,v,o}
    \fmfv{d.f=empty,d.shape=circle,d.size=3.5mm}{v}
\end{fmfgraph*}}
$
which is the sum of $\sigma^{(2)}$ and all of its one-particle irreducible (1PI) corrections. Note that $\sigma_{R,ij}^{(2)}$ has external legs corresponding to both $\varphi$ and $\tilde \varphi$ (similarly to $\D_{ij}$).
It is therefore more appropriate to consider the physical fields and the response fields on an equal footing, as indicated by the double legs representing $\psi_i$. We denote
$\D
=
\parbox{10mm}{
\centering
\begin{fmfgraph*}(10,5)
    \setval
    \fmfleft{i}
    \fmfright{o}
    \fmf{Double}{i,o}
\end{fmfgraph*}
}
$,
and the corresponding renormalized propagator 
$\D_R
=
\parbox{10mm}{
\centering
\begin{fmfgraph*}(10,3.5)
    \setval
    \fmfleft{i}
    \fmfright{o}
    \fmf{Double}{i,v,o}
    \fmfv{d.f=hatched,d.shape=circle,d.size=3.5mm}{v}
\end{fmfgraph*}
}
$.
We may now write out the interacting version of Eq.~\eqref{eq: diagrams} by simply substituting the propagators and vertices with their renormalized counterparts. This yields a result similar to the tree-level summation, namely
\begin{widetext}
\begin{align}\label{eq: Delta R}
    \Delta_{R,ij}^{(1)}& = 
    -\,
    \parbox{10mm}{
    \begin{fmfgraph*}(10,5)
        \setval
        \fmfleft{i}
        \fmfright{o}
        \fmf{Double}{i,v,o}
        \fmfv{d.f=hatched,d.shape=circle,d.size=3.5mm}{v}
    \end{fmfgraph*}
    }
    +
    \Big(
    \parbox{10mm}{
    \begin{fmfgraph*}(10,5)
        \setval
        \fmfleft{i}
        \fmfright{o}
        \fmf{Double}{i,v,o}
        \fmfv{d.f=hatched,d.shape=circle,d.size=3.5mm}{v}
    \end{fmfgraph*}
    }
    +
    \parbox{18mm}{
    \begin{fmfgraph*}(18,5)
        \setval
        \fmfleft{i}
        \fmfright{o}
        \fmf{Double}{i,c1,v,c2,o}
        \fmfv{d.f=empty,d.shape=circle,d.size=3.5mm}{v}
        \fmfv{d.f=hatched,d.shape=circle,d.size=3.5mm}{c1}
        \fmfv{d.f=hatched,d.shape=circle,d.size=3.5mm}{c2}
    \end{fmfgraph*}
    }
    +
    \parbox{26mm}{
    \begin{fmfgraph*}(26,5)
        \setval
        \fmfleft{i}
        \fmfright{o}
        \fmf{Double}{i,c1,v1,c2,v2,c3,o}
        \fmfv{d.f=empty,d.shape=circle,d.size=3.5mm}{v1}
        \fmfv{d.f=empty,d.shape=circle,d.size=3.5mm}{v2}
        \fmfv{d.f=hatched,d.shape=circle,d.size=3.5mm}{c1}
        \fmfv{d.f=hatched,d.shape=circle,d.size=3.5mm}{c2}
        \fmfv{d.f=hatched,d.shape=circle,d.size=3.5mm}{c3}
    \end{fmfgraph*}
    }
    + \cdots
    \Big)
    \Big(
    1+
    \,\,\,
    \parbox{5mm}{
    \begin{fmfgraph*}(5,5)
        \setval
        \fmfleft{i}
        \fmfright{o}
        \fmf{Double,left}{i,o}
        \fmf{Double,left}{o,i}
        \fmfv{d.f=empty,d.shape=circle,d.size=3.5mm}{i}
    \end{fmfgraph*}
    }
    +
    \frac{1}{2!}
    \Big[
    \parbox{8mm}{
    \,
    \begin{fmfgraph*}(5,5)
        \setval
        \fmfleft{i}
        \fmfright{o}
        \fmf{Double,left}{i,o,i}
        \fmfv{d.f=empty,d.shape=circle,d.size=3.5mm}{i}
        \fmfv{d.f=empty,d.shape=circle,d.size=3.5mm}{o}
    \end{fmfgraph*}}
    \,+2\hspace{3mm}
    \parbox{7mm}{
    ${
    \begin{fmfgraph*}(5,5)
        \setval
        \fmfleft{i}
        \fmfright{o}
        \fmf{Double,left}{i,o}
        \fmf{Double,left}{o,i}
        \fmfv{d.f=empty,d.shape=circle,d.size=3.5mm}{i}
    \end{fmfgraph*}
    }^2$
    }
    \Big]
    +\cdots
    \Big),
\end{align}
\end{widetext}
where the expression in the second parentheses adds up to unity
The details are presented in Appendix A, while the calculation of $\sigma^{(2)}_{R}$ is shown in Appendix B.
The calculation of $\Delta^{(2)}$ is similar, though not without subtleties, and can be found in details in Appendix C.
Similarly to the case at tree-level, $\Delta^{(1)}$ and $\Delta^{(2)}$ are written in terms of re-summed entropy operators, $\Sigma_R^{(2)}$ and $\Sigma_R^{(4)}$, respectively.
However, these now contain the perturbative corrections to the entropy consumption in the form of regularized quantities: $\delta r$, $\delta \alpha_0$, and the divergent integral $\integral$.

\textit{Discussion---}%
Using the one-loop calculations, we observe that by defining $\Sigma_R \equiv {\Sigma^{(2)}_R} +  {\Sigma^{(4)}_{R}}$ we obtain a result that is independent of $\integral$, with only the lower-right quadrant being non-zero. This calculation yields the following FDRs, verified at one-loop order:
\begin{align}
    \label{eq: GFDT R 1}
    C_{R+} - C_{R-} &= C_{R-} \Sigma_{R+} C_{R-},\\
    \label{eq: GFDT R 2}
    G_{R+} - G_{R-} - \frac{ i\omega }{D q^2} C_{R+} &= C_{R-} \Sigma_{R+} G_{R-},\\
    \label{eq: GFDT R 3}
    \frac{ i\omega }{D q^2}\left[G_{R+} - G_{R+}^\dagger - \frac{ i\omega }{D q^2} C_{R-} \right] &= G_{R-}^\dagger \Sigma_{R+} G_{R-}.
\end{align}

We may utilize these expressions to eliminate $\Sigma$.
From Eq.~\eqref{eq: GFDT R 1}, we obtain $\Sigma_{R+} = C_{R-}^{-1} \left( C_{R+}C_{R-}^{-1} - \one \right)$, which can be inserted into Eq.~\eqref{eq: GFDT R 2} to yield Eq.~\eqref{eq: FDR}. Similarly, inserting this form into Eq.~\eqref{eq: GFDT R 3} gives
\begin{align}
2i\omega \left( G - G^\dagger - \frac{i\omega}{q^2 D} C\right) 
 = \left| { {G^*}^{-1}G }\right|^2-I,
\end{align}
where we employ the notation $|A|^2 \equiv A A^\dagger$ and drop the subscripts for notational convenience.
We observe that the same motif appears in both of these equations: the deviation of ${C^*}^{-1}C$ and $|{G^*}^{-1}G|^2$ from the identity captures the deviation away from the equilibrium FDT.
These identities are as stringent as their equilibrium counterparts, in the following sense: given a measurement of the physical correlation function $C_\text{phy}$ of some system, one can infer the properties of $G_\text{phy}$, or $\chi_\text{phy}$, directly.

As discussed in Ref. \cite{johnsrudFluctuationDissipationRelations2025a}, $\Delta$ (and thus $\Sigma$) can be related to total entropy $\E{S}$ using the field-theoretic Harada-Sasa relation~\cite{nardiniEntropyProductionField2017}, by applying the functional trace $\Tr A \equiv L^d T \int_{\bm q, \omega } A_{aa}(\bm q, \omega)$ where $L$ is the system size and $T$ is the total observation time. 
We find \cite{johnsrudFluctuationDissipationRelations2025a}
\begin{align}
    \E{-S}
    = \Tr q^2 D {G^*}^\dagger \Sigma G^*
    = \Tr \Sigma C^*,
\end{align}
where we have used Eq. \eqref{eq: CGG causality} and the cyclic property of the trace.
Since $\Sigma C^* = {C^*}^{-1}C - I$, we find that the motif discussed above is again observed. We can compare this result to the entropy obtained for the linear theory, given by the bubble diagram 
$ \E{- S_0}_0 = 
\parbox{3mm}{
\centering
\begin{fmfgraph*}(3,3)
    \setval
    \fmfleft{i}
    \fmfright{o}
    \fmf{plain,left}{i,o}
    \fmf{plain,left}{o,i}
    \fmfdot{o}
\end{fmfgraph*}}
=
\Tr \sigma C$ (see Ref. \cite{supp}).
This illustrates that, indeed, $\Sigma_R$ is a renormalized version of the two-point entropy consumption vertex $\sigma$. We note further that $\Sigma C^*$ is precisely the operator that appears in the diagram $d^{(2)}$ in Appendix B.

The procedure presented in this paper can be applied to any out-of-equilibrium system that satisfies the assumptions of \cite{johnsrudFluctuationDissipationRelations2025a} (mainly time and space translational invariance). The recipe is as follows: (i) Define the time-reversal transformation $\T$. As discussed in~\cite{johnsrudFluctuationDissipationRelations2025a}, the lack of a $\T$-symmetry can introduce some freedom in choosing $\T$. A fortuitous choice may even yield exact results at this step, such as the case of models with odd mobility \cite{johnsrudFluctuationDissipationRelations2025a}. (ii) Calculate the entropy $S = \T A - A$, and use it to write down the FDRs [Eq.~\eqref{eq: GFDT matrix}], and extract the entropy consumption vertices $\sigma$. Depending on the theory, these may have two or more legs. (iii) Perform the diagrammatic expansion and calculate the diagrams. At this stage there might be simplifications for any given theory, such as the following observation in the case of NRCH: the fact that only the lower-right quadrant of $\Sigma_R$ is non-zero allowed us to derive Eq.~\eqref{eq: FDR}. Another example can be observed for the case of KPZ: 
if the dynamics involves only one field, then $C = C^* $ and $G^* = G^\dagger$, thereby connecting the FDRs and greatly reducing the calculations needed.

To summarize, we have introduced a framework to quantify deviations from the equilibrium fluctuation dissipation relations in active field theories, in the form of a diagrammatic formalism that allows for perturbative calculations of the observables of interest.
With this, we are able to perturbatively write down identities relating the correlations and the response functions, as well as the entropy production.
The calculations performed here demonstrate the strength of the developed framework using the specific example of the NRCH model, and gives important insight into the particular non-equilibrium character of NRCH. In the accompanying paper, we sketch how similar calculations can be performed for other active field theories \cite{johnsrudFluctuationDissipationRelations2025a} as well as other non-equilibrium field theories such as the KPZ equation \cite{kardarDynamicScalingGrowing1986}.

\begin{acknowledgements}
We acknowledge support from the Max Planck School Matter to Life and the MaxSynBio Consortium which are jointly funded by the Federal Ministry of Education and Research (BMBF) of Germany and the Max Planck Society.    
\end{acknowledgements}

\appendix

\begin{widetext}

\begin{center}
\textbf{Appendix}
\end{center}


\textit{Appendix A: Loop expansion of $\Delta^{(1)}$---}%
Calculations in nonlinear field theories are done by expressing expectation values of the full interacting field theory, as perturbation series in the free theory denoted by a zero subscript, $\E{\Oh} = \E{\Oh e^{-A_I}}_0$. Here, $A_I$ is the part of the response-field action containing the interaction-vertex. Thus, $\Delta_{R,ij}^{(1)} \equiv \E{\psi_i\psi_j\left(e^{-S_2} - 1\right)e^{-A_I}}_0 $ is an expansion both in the number of loops and powers of $\alpha_0$, leading to a large set of diagrams. 

Consider the diagrams that this expansion will give rise to. First, we see that where there is a propagator in Eq.~\eqref{eq: linear expansion}, there will now be a renormalized propagator. In addition, there will be new loops in the bubbles, and diagrams similar to those that renormalize $C$ and $G$, but containing one or more entropy-consumption vertices. As an illustration, we write the first few diagrams in the expansion of $\Delta_{ab}^{(1)}$,
\begin{align}\nonumber
    \Delta_{R,ab}^{(1)}& = 
    \parbox{18mm}{
    \centering
    \begin{fmfgraph*}(18,6)
        \setval
        \fmfleft{i}
        \fmfright{o}
        \fmf{plain}{i,v1,v,v2,o}
        \fmfv{d.f=hatched,d.shape=circle,d.size=3.5mm}{v1}
        \fmfv{d.f=hatched,d.shape=circle,d.size=3.5mm}{v2}
        \fmfdot{v}
    \end{fmfgraph*}
    }
    +
    \parbox{28mm}{
    \centering
    \begin{fmfgraph*}(4,4)
        \setval
        \fmfleft{i}
        \fmfright{o}
        \fmf{plain,left}{i,o}
        \fmf{plain,left}{o,i}
        \fmfdot{o}
    \end{fmfgraph*}
    +
    \begin{fmfgraph*}(8,4)
        \setval
        \fmfleft{i}
        \fmfright{o}
        \fmf{plain,left}{c,o,c}
        \fmf{plain,left}{c,i}
        \fmf{response,left}{i,c}
        \fmfdot{i}
    \end{fmfgraph*}
    +\dots\\
    \begin{fmfgraph*}(26,5)
        \setval
        \fmfleft{i}
        \fmfright{o}
        \fmf{plain}{i,v,o}
        \fmfv{d.f=hatched,d.shape=circle,d.size=3.5mm}{v}
    \end{fmfgraph*}
    }
    + 
    \parbox{22mm}{
    \centering
    \begin{fmfgraph*}(22,6)
        \setval
        \fmfleft{i}
        \fmfright{o}
        \fmf{plain}{i,v1,v,v2,v3,v4,o}
        \fmfv{d.f=hatched,d.shape=circle,d.size=3.5mm}{v1}
        \fmfv{d.f=hatched,d.shape=circle,d.size=3.5mm}{v2}
        \fmfv{d.f=hatched,d.shape=circle,d.size=3.5mm}{v4}
        \fmfdot{v}
        \fmfdot{v3}
    \end{fmfgraph*}
    }
    +
    \parbox{26mm}{
    \centering
    \begin{fmfgraph*}(4,4)
        \setval
        \fmfleft{i}
        \fmfright{o}
        \fmf{plain,left}{i,o}
        \fmf{plain,left}{o,i}
        \fmfdot{o}
    \end{fmfgraph*}
    +
    \centering
    \begin{fmfgraph*}(8,4)
        \setval
        \fmfleft{i}
        \fmfright{o}
        \fmf{plain,left}{c,o,c}
        \fmf{plain,left}{c,i}
        \fmf{response,left}{i,c}
        \fmfdot{i}
    \end{fmfgraph*}
    +\dots\\
    \begin{fmfgraph*}(26,5)
        \setval
        \fmfleft{i}
        \fmfright{o}
        \fmf{plain}{i,c1,c2,c3,o}
        \fmfv{d.f=hatched,d.shape=circle,d.size=3.5mm}{c1}
        \fmfv{d.f=hatched,d.shape=circle,d.size=3.5mm}{c3}
        \fmfdot{c2}
    \end{fmfgraph*}
    }
    +
    \parbox{35mm}{
    \centering
    \begin{fmfgraph*}(3,3)
        \setval
        \fmfleft{i}
        \fmfright{o}
        \fmf{plain,left}{i,o,i}
        \fmfdot{i}
        \fmfdot{o}
    \end{fmfgraph*}
    +
    $
    \begin{fmfgraph*}(3,3)
        \setval
        \fmfleft{i}
        \fmfright{o}
        \fmf{plain,left}{i,o}
        \fmf{plain,left}{o,i}
        \fmfdot{o}
    \end{fmfgraph*}^2
    $
    +
    \begin{fmfgraph*}(6,3)
        \setval
        \fmfleft{o}
        \fmfright{i}
        \fmf{plain,left}{o,c,i,c}
        \fmf{wiggly,left}{c,o}
        \fmfforce{.75w,1h}{a}
        \fmfforce{.75w,0h}{b}
        \fmfdot{a}
        \fmfdot{b}
    \end{fmfgraph*}
    + \dots\\
    \begin{fmfgraph*}(35,5)
        \setval
        \fmfleft{i}
        \fmfright{o}
        \fmf{plain}{i,v,o}
        \fmfv{d.f=hatched,d.shape=circle,d.size=3.5mm}{v}
    \end{fmfgraph*}
    }
    + \cdots\\\nonumber
    & \quad +
    \parbox{16mm}{
    \centering
    \begin{fmfgraph*}(16,10)
        \setval
        \fmfleft{i}
        \fmfright{o}
        \fmftop{t}
        \fmf{plain}{i,i1}
        \fmf{plain}{i1,c}
        \fmf{plain}{o,o1}
        \fmf{wiggly}{o1,c}
        \fmffreeze
        \fmf{plain,left}{c,t}
        \fmf{plain,left}{t,c}
        \fmfdot{t}
    \end{fmfgraph*}
    }
    +
    \parbox{16mm}{
    \centering
    \begin{fmfgraph*}(16,10)
        \setval
        \fmfleft{i}
        \fmfright{o}
        \fmftop{t}
        \fmf{plain}{i,c}
        \fmf{plain}{o,c}
        \fmffreeze
        \fmf{plain,left=.4}{c,a,t,b}
        \fmf{wiggly,left=.4}{b,c}
        \fmfforce{.34w, .75h}{a}
        \fmfforce{.66w, .75h}{b}
        \fmfdot{t}
    \end{fmfgraph*} 
    }
    +
    \parbox{16mm}{
    \centering
    \begin{fmfgraph*}(16,10)
        \setval
        \fmfleft{i}
        \fmfright{o}
        \fmf{plain,tension=2}{i,i1}
        \fmf{wiggly,tension=2}{i1,c1}
        \fmf{plain,tension=2}{o,o1}
        \fmf{wiggly,tension=2}{o1,c2}
        \fmf{plain,tension=1/6}{c1,c3}
        \fmf{plain,tension=1/6}{c3,c2}
        \fmf{plain,left,tension=1/3}{c1,c2}
        \fmf{plain,right,tension=1/3}{c1,c2}
        \fmfdot{c3}
    \end{fmfgraph*}
    }
    +
    \parbox{16mm}{
    \centering
    \begin{fmfgraph*}(16,6)
        \setval
        \fmfleft{i}
        \fmfright{o}
        \fmftop{t}
        \fmf{plain}{i,i1}
        \fmf{wiggly}{i1,c1}
        \fmf{plain}{o,o1}
        \fmf{plain}{o1,c2}
        \fmf{plain,tension=.8}{c1,c3}
        \fmf{wiggly,tension=.8}{c3,c2}
        \fmffreeze
        \fmf{plain,left=.45,tension=1/2}{c1,t}
        \fmf{plain,left=.45,tension=1/2}{t,c2}
        \fmf{plain,right}{c1,c2}
        \fmfdot{t}
    \end{fmfgraph*}
    }
    +
    \parbox{16mm}{
    \centering
    \begin{fmfgraph*}(16,6)
        \setval
        \fmfleft{i}
        \fmfright{o}
        \fmftop{t}
        \fmf{plain}{i,i1}
        \fmf{plain}{i1,c1}
        \fmf{plain}{o,o1}
        \fmf{plain}{o1,c2}
        \fmf{plain,tension=1.6}{c1,c3}
        \fmf{plain,tension=1.6}{c3,c4}
        \fmf{wiggly,tension=.8}{c4,c2}
        \fmfdot{c3}
        \fmffreeze
        \fmf{wiggly,left=.45,tension=1/2}{c1,t}
        \fmf{plain,left=.45,tension=1/2}{t,c2}
        \fmf{plain,right}{c1,c2}
    \end{fmfgraph*}
    }
    +
    \parbox{16mm}{
    \centering
    \begin{fmfgraph*}(16,6)
        \setval
        \fmfleft{i}
        \fmfright{o}
        \fmftop{t}
        \fmfbottom{b}
        \fmf{plain}{i,i1}
        \fmf{plain}{i1,c1}
        \fmf{plain}{o,o1}
        \fmf{plain}{o1,c2}
        \fmf{plain,tension=1.6}{c1,c3}
        \fmf{plain,tension=1.6}{c3,c4}
        \fmf{wiggly,tension=.8}{c4,c2}
        \fmfdot{b}
        \fmffreeze
        \fmf{wiggly,left=.45,tension=1/2}{c1,t}
        \fmf{plain,left=.45,tension=1/2}{t,c2}
        \fmf{plain,right}{c1,c2}
    \end{fmfgraph*}
    }
    + \dots\\
    &\quad +
    \parbox{16mm}{
    \centering
    \begin{fmfgraph*}(16,10)
        \setval
        \fmfleft{i}
        \fmfright{o}
        \fmftop{t}
        \fmf{plain}{i,i1}
        \fmf{plain}{i1,c}
        \fmf{plain}{o,o1}
        \fmf{wiggly}{o1,c}
        \fmffreeze
        \fmf{plain,left=.45}{c,c1,t,c2,c}
        \fmfdot{c1}
        \fmfdot{c2}
        \fmfforce{.35w, .75h}{c1}
        \fmfforce{.65w, .75h}{c2}
    \end{fmfgraph*}
    }
    +
    \parbox{16mm}{
    \centering
    \begin{fmfgraph*}(16,10)
        \setval
        \fmfleft{i}
        \fmfright{o}
        \fmftop{t}
        \fmf{plain}{i,c}
        \fmf{plain}{o,c}
        \fmffreeze
        \fmf{plain,left=.4}{c,a,t,b}
        \fmf{wiggly,left=.4}{b,c}
        \fmfforce{.34w, .75h}{a}
        \fmfforce{.66w, .75h}{b}
        \fmfdot{t}
        \fmfdot{a}
    \end{fmfgraph*}
    }
    +
    \parbox{16mm}{
    \centering
    \begin{fmfgraph*}(16,6.5)
        \setval
        \fmfleft{i}
        \fmfright{o}
        \fmftop{t}
        \fmfbottom{b}
        \fmf{plain,tension=2}{i,i1}
        \fmf{wiggly,tension=2}{i1,c1}
        \fmf{plain}{c2,o}
        \fmf{plain,tension=1/6}{c1,c3}
        \fmf{plain,tension=1/6}{c3,c2}
        \fmf{plain,left=.45,tension=2/3}{c1,t}
        \fmf{plain,left=.45,tension=2/3}{t,c2}
        \fmf{plain,right=.45,tension=2/3}{c1,b}
        \fmf{wiggly,right=.45,tension=2/3}{b,c2}
        \fmfdot{c3}
        \fmfdot{t}
    \end{fmfgraph*}
    } 
    +
    \parbox{20mm}{
    \centering
    \begin{fmfgraph*}(20,10)
        \setval
        \fmfleft{i}
        \fmfright{o}
        \fmf{plain}{i,c1}
        \fmf{wiggly,tension=2}{c1,c2}
        \fmf{plain,tension=2}{c2,c3}
        \fmf{wiggly,tension=2}{c3,c4}
        \fmf{plain,tension=2}{c4,o}
        \fmffreeze
        \fmfforce{.33w,1h}{t1}
        \fmfforce{.67w,1h}{t2}
        \fmf{plain,left}{c1,t1}
        \fmf{plain,left}{t1,c1}
        \fmf{plain,left}{c3,t2}
        \fmf{plain,left}{t2,c3}
        \fmfdot{t1}
    \end{fmfgraph*}
    }
    +
    \parbox{20mm}{
    \centering
    \begin{fmfgraph*}(20,10)
        \setval
        \fmfleft{i}
        \fmfright{o}
        \fmf{plain}{i,c1}
        \fmf{wiggly,tension=2}{c1,c2}
        \fmf{plain,tension=2}{c2,c3}
        \fmf{wiggly,tension=2}{c3,c4}
        \fmf{plain,tension=2}{c4,o}
        \fmffreeze
        \fmfforce{.33w,1h}{t1}
        \fmfforce{.67w,1h}{t2}
        \fmf{plain,left}{c1,t1}
        \fmf{plain,left}{t1,c1}
        \fmf{plain,left}{c3,t2}
        \fmf{plain,left}{t2,c3}
        \fmfdot{t1}
        \fmfdot{t2}
    \end{fmfgraph*}
    }
    + 
    \parbox{20mm}{
    \centering
    \begin{fmfgraph*}(20,10)
        \setval
        \fmfleft{i}
        \fmfright{o}
        \fmf{plain}{i,c1}
        \fmf{wiggly,tension=2}{c1,c2}
        \fmf{plain,tension=2}{c2,c3}
        \fmf{wiggly,tension=2}{c3,c4}
        \fmf{plain,tension=2}{c4,o}
        \fmffreeze
        \fmfforce{.33w,1h}{t1}
        \fmfforce{.21w,.75h}{a1}
        \fmfforce{.45w,.75h}{a2}
        \fmfforce{.67w,1h}{t2}
        \fmf{plain,left=.45}{c1,a1,t1,a2,c1}
        \fmf{plain,left}{c3,t2}
        \fmf{plain,left}{t2,c3}
        \fmfdot{a1}
        \fmfdot{a2}
    \end{fmfgraph*}
    }
    + 
    \cdots. 
\end{align}
Here, we have only included enough terms to illustrate how we can perform a resummation of the expansion. In fact, we can capture all the remaining new terms with the definition of the renormalized entropy consumption operator, 
\begin{align}\label{eq: sigma R}
    \sigma_{R,ij}^{(2)}
    = 
    \parbox{10mm}{
    \centering
    \begin{fmfgraph*}(10,6)
        \setval
        \fmfleft{i}
        \fmfright{o}
        \fmf{plain}{i,v,o}
        \fmfdot{v}
    \end{fmfgraph*}
    }
    +
    \parbox{12mm}{
    \centering
    \begin{fmfgraph*}(12,10)
        \setval
        \fmfleft{i}
        \fmfright{o}
        \fmftop{t}
        \fmf{plain}{i,c}
        \fmf{wiggly}{o,c}
        \fmffreeze
        \fmf{plain,left}{c,t}
        \fmf{plain,left}{t,c}
        \fmfdot{t}
    \end{fmfgraph*}
    }
    +
    \parbox{12mm}{
    \centering
    \begin{fmfgraph*}(12,10)
        \setval
        \fmfleft{i}
        \fmfright{o}
        \fmftop{t}
        \fmf{plain}{i,c}
        \fmf{plain}{o,c}
        \fmffreeze
        \fmf{plain,left=.4}{c,a,t,b}
        \fmf{wiggly,left=.4}{b,c}
        \fmfforce{.3w, .75h}{a}
        \fmfforce{.7w, .75h}{b}
        \fmfdot{t}
    \end{fmfgraph*} 
    }
    +
    \parbox{12mm}{
    \centering
    \begin{fmfgraph*}(12,10)
        \setval
        \fmfleft{i}
        \fmfright{o}
        \fmf{wiggly}{i,c1}
        \fmf{wiggly}{o,c2}
        \fmf{plain,tension=1}{c1,c3}
        \fmf{plain,tension=1}{c3,c2}
        \fmffreeze
        \fmf{plain,left,tension=1/3}{c1,c2}
        \fmf{plain,right,tension=1/3}{c1,c2}
        \fmfdot{c3}
    \end{fmfgraph*}
    }
    +
    \parbox{12mm}{
    \centering
    \begin{fmfgraph*}(12,6)
        \setval
        \fmfleft{i}
        \fmfright{o}
        \fmftop{t}
        \fmf{plain}{i,c1}
        \fmf{plain}{o,c2}
        \fmf{plain,tension=.8}{c1,c3}
        \fmf{wiggly,tension=.8}{c3,c2}
        \fmffreeze
        \fmf{plain,left=.4}{c1,t}
        \fmf{plain,left=.4}{t,c2}
        \fmf{plain,right}{c1,c2}
        \fmfdot{t}
    \end{fmfgraph*}
    }
    + \cdots
    +
    \parbox{12mm}{
    \centering
    \begin{fmfgraph*}(12,10)
        \setval
        \fmfleft{i}
        \fmfright{o}
        \fmftop{t}
        \fmf{plain}{i,c}
        \fmf{wiggly}{o,c}
        \fmffreeze
        \fmf{plain,left=.45}{c,c1,t,c2,c}
        \fmfdot{c1}
        \fmfdot{c2}
        \fmfforce{.3w, .75h}{c1}
        \fmfforce{.7w, .75h}{c2}
    \end{fmfgraph*}
    }
    +
    \parbox{12mm}{
    \centering
    \begin{fmfgraph*}(12,10)
        \setval
        \fmfleft{i}
        \fmfright{o}
        \fmftop{t}
        \fmf{plain}{i,c}
        \fmf{plain}{o,c}
        \fmffreeze
        \fmf{plain,left=.4}{c,a,t,b}
        \fmf{wiggly,left=.4}{b,c}
        \fmfforce{.3w, .75h}{a}
        \fmfforce{.7w, .75h}{b}
        \fmfdot{t}
        \fmfdot{a}
    \end{fmfgraph*}
    }
    +
    \parbox{12mm}{
    \centering
    \begin{fmfgraph*}(12,5)
        \setval
        \fmfleft{i}
        \fmfright{o}
        \fmftop{t}
        \fmfbottom{b}
        \fmf{wiggly,tension=1.5}{i,c1}
        \fmf{plain,tension=1.5}{c2,o}
        \fmf{plain,tension=2}{c1,c3}
        \fmf{plain,tension=2}{c3,c2}
        \fmffreeze
        \fmf{plain,left=.45,tension=2/3}{c1,t}
        \fmf{plain,left=.45,tension=2/3}{t,c2}
        \fmf{plain,right=.45,tension=2/3}{c1,b}
        \fmf{wiggly,right=.45,tension=2/3}{b,c2}
        \fmfdot{c3}
        \fmfdot{t}
    \end{fmfgraph*}
    } 
    + \cdots.
\end{align}
Notice that these diagrams are ``amputated''. To obtain the diagrams of $\Delta_R$, we attach propagators to the end of each side of the diagrams.
For example,
$
\parbox{12mm}{
\centering
\begin{fmfgraph*}(12,2.5)
    \setval
    \fmfforce{0w,0h}{i}
    \fmfforce{1w,0h}{o}
    \fmftop{t}
    \fmf{plain}{i,i1}
    \fmf{plain}{i1,c}
    \fmf{plain}{o,o1}
    \fmf{wiggly}{o1,c}
    \fmffreeze
    \fmf{plain,left}{c,t}
    \fmf{plain,left}{t,c}
    \fmfdot{t}
\end{fmfgraph*}
}
=
C\,
\parbox{8mm}{
\centering
\begin{fmfgraph*}(8,2.5)
    \setval
    \fmfforce{0w,0h}{i}
    \fmfforce{1w,0h}{o}
    \fmftop{t}
    \fmf{plain}{i,c}
    \fmf{wiggly}{o,c}
    \fmffreeze
    \fmf{plain,left}{c,t}
    \fmf{plain,left}{t,c}
    \fmfdot{t}
\end{fmfgraph*}
}
\,
G^\dagger.
$
Notice here that the external legs correspond to both $\varphi$ and $i \tilde \varphi$, which is why we have given the object $ij$-indices.
One can see that the definition given above directly captures all diagrams in the second line, and the three first ones in the third line.
The three last terms in the third line are then constructed by chaining together $C_R$, $G_R$, and $\sigma_R^{(2)}$, in the same way as was done with $C$, $G$, and $\sigma$ before. 
These three particular diagrams are captured by $C_R\sigma_R^{(2)} G_R \sigma_R^{(2)} G_R$, leading again to a Dyson series. 
The second and third lines are thus captured by the diagram in the first parenthesis of Eq.~\eqref{eq: Delta R}.

In fact, this definition also captures all the bubble diagrams.
One might expect to have bubble diagrams that connect entropy vertices $\sigma_R^{(2)}$ with the renormalized propagators $\D_{R}$. However, by cutting one of the internal legs of $\D_R$, we notice that this diagram corresponds to a bubble with only $\sigma_R^{(2)}$. 
In diagramatic terms,
$
\text{\ScissorRight}
\,\,\,
\parbox{5mm}{
\begin{fmfgraph*}(4,4)
    \setval
    \fmfleft{i}
    \fmfright{o}
    \fmf{Double,left}{i,o}
    \fmf{Double,left}{o,i}
    \fmfv{d.f=hatched,d.shape=circle,d.size=3mm}{i}
    \fmfv{d.f=empty,d.shape=circle,d.size=3mm}{o}
\end{fmfgraph*}
}\,
\rightarrow
\parbox{8mm}{
\begin{fmfgraph*}(8,3)
    \setval
    \fmfforce{0w,0h}{i}
    \fmfforce{1w,0h}{o} 
    \fmftop{t}
    \fmf{Double}{i,c,o}
    \fmffreeze
    \fmf{Double,left=1.1}{c,t,c}
    \fmfv{d.f=hatched,d.shape=circle,d.size=2.5mm}{c}
    \fmfv{d.f=empty,d.shape=circle,d.size=2.5mm}{t}
\end{fmfgraph*}
}
\in
\parbox{8mm}{
\begin{fmfgraph*}(8,5)
    \setval
    \fmfleft{i}
    \fmfright{o}
    \fmf{Double}{i,c,o}
    \fmfv{d.f=empty,d.shape=circle,d.size=3.5mm}{c}
\end{fmfgraph*}
}
$.
As a specific example, take
$
\text{\ScissorRight}
\,
\parbox{6mm}{
    \centering
    \begin{fmfgraph*}(6,4)
        \setval
        \fmfleft{o}
        \fmfright{i}
        \fmf{plain,left}{o,c,i,c}
        \fmf{wiggly,left}{c,o}
        \fmfdot{i}
    \end{fmfgraph*}
}\,
\longrightarrow
\parbox{8mm}{
\centering
\begin{fmfgraph*}(8,3)
    \setval
    \fmftop{t}
    \fmfforce{0w,0h}{i}
    \fmfforce{1w,0h}{o} 
    \fmf{plain}{i,c}
    \fmf{wiggly}{o,c}
    \fmffreeze
    \fmf{plain,left}{c,t,c}
    \fmfdot{t}
\end{fmfgraph*} 
}
\in
\parbox{8mm}{
\begin{fmfgraph*}(8,4)
    \setval
    \fmfleft{i}
    \fmfright{o}
    \fmf{Double}{i,c,o}
    \fmfv{d.f=empty,d.shape=circle,d.size=3.5mm}{c}
\end{fmfgraph*}
}
$.
It would therefore be double-counting to include these bubbles.
With these definitions, the combinatorics are exactly the same as in the linear case, as detailed in~\cite{supp}, and again, the bubble diagrams factor out.
The case for the other quadrants of $\Delta_{ij}^{(1)}$ are similar, only with different external legs, so the full expansion is captured by Eq.~\eqref{eq: Delta R}.


\textit{Appendix B: One loop diagrams of $\sigma_R^{(2)}$---}%
The leading order in $\alpha_0$ corrections to the entropy consumption is
\begin{align}
    \sigma_{R, ij}^{(2)}
    =
    \begin{pmatrix}
        0 &\quad& {
            \parbox{10mm}{
            \centering
            \begin{fmfgraph*}(10,8)
                \setval
                \fmfleft{i}
                \fmfright{o}
                \fmftop{t}
                \fmf{plain}{i,c}
                \fmf{wiggly}{o,c}
                \fmffreeze
                \fmf{plain,left}{c,t}
                \fmf{plain,left}{t,c}
                \fmfdot{t}
            \end{fmfgraph*}
            }
            }^\dagger \\[6mm]
            \parbox{10mm}{
            \centering
            \begin{fmfgraph*}(10,8)
                \setval
                \fmfleft{i}
                \fmfright{o}
                \fmftop{t}
                \fmf{plain}{i,c}
                \fmf{wiggly}{o,c}
                \fmffreeze
                \fmf{plain,left}{c,t}
                \fmf{plain,left}{t,c}
                \fmfdot{t}
            \end{fmfgraph*}
            }
        &\quad&
        \parbox{10mm}{
        \centering
        \begin{fmfgraph*}(10,6)
            \setval
            \fmfleft{i}
            \fmfright{o}
            \fmf{plain}{i,v,o}
            \fmfdot{v}
        \end{fmfgraph*}
        }
        +
        \parbox{10mm}{
        \centering
        \begin{fmfgraph*}(10,8)
            \setval
            \fmfleft{i}
            \fmfright{o}
            \fmftop{t}
            \fmf{plain}{i,c}
            \fmf{plain}{o,c}
            \fmffreeze
            \fmf{plain,left=.4}{c,a,t,b}
            \fmf{wiggly,left=.4}{b,c}
            \fmfforce{.3w, .75h}{a}
            \fmfforce{.7w, .75h}{b}
            \fmfdot{t}
        \end{fmfgraph*} 
        }
    \end{pmatrix}.
\end{align}
These diagrams are the same as those renormalizing $G$ and $C$, detailed in~\cite{supp}, but with the substitution $G\rightarrow C\sigma^{(2)} G$ and $C\rightarrow C\sigma^{(2)} C$.
Calculating these integrals directly gives a divergence that does not seemingly match $\delta r$. However, they represent only the first set of one-loop corrections in an infinite series. For consistency, one has to take into account one-loop corrections to all orders in $\alpha_0$, which results in an exact cancellation, namely
\begin{align}
    d^{(1)}_{ab}
    \equiv
    \parbox{12mm}{
    \centering
    \begin{fmfgraph*}(12,10)
        \setval
        \fmfleft{i}
        \fmfright{o}
        \fmftop{t}
        \fmf{plain}{i,c}
        \fmf{wiggly}{o,c}
        \fmffreeze
        \fmf{plain,left}{c,t}
        \fmf{plain,left}{t,c}
        \fmfv{d.f=empty,d.shape=triangle,d.size=2mm}{t}
    \end{fmfgraph*}
    }
    & \equiv
    \parbox{12mm}{
    \centering
    \begin{fmfgraph*}(12,10)
        \setval
        \fmfleft{i}
        \fmfright{o}
        \fmftop{t}
        \fmf{plain}{i,c}
        \fmf{wiggly}{o,c}
        \fmffreeze
        \fmf{plain,left}{c,t}
        \fmf{plain,left}{t,c}
        \fmfdot{t}
    \end{fmfgraph*}
    }
    +
    \parbox{12mm}{
    \centering
    \begin{fmfgraph*}(12,10)
        \setval
        \fmfleft{i}
        \fmfright{o}
        \fmftop{t}
        \fmf{plain}{i,c}
        \fmf{wiggly}{o,c}
        \fmffreeze
        \fmf{plain,left=.4}{c,a,t,b}
        \fmf{plain,left=.4}{b,c}
        \fmfforce{.3w, .75h}{a}
        \fmfforce{.7w, .75h}{b}
        \fmfdot{b}
        \fmfdot{a}
    \end{fmfgraph*}
    }
    +
    \parbox{12mm}{
    \centering
    \begin{fmfgraph*}(12,10)
        \setval
        \fmfleft{i}
        \fmfright{o}
        \fmftop{t}
        \fmf{plain}{i,c}
        \fmf{wiggly}{o,c}
        \fmffreeze
        \fmf{plain,left=.4}{c,a,t,b}
        \fmf{plain,left=.4}{b,c}
        \fmfforce{.3w, .75h}{a}
        \fmfforce{.7w, .75h}{b}
        \fmfdot{b}
        \fmfdot{a}
        \fmfdot{t}
    \end{fmfgraph*}
    }
    +\cdots
    = -3 q^2 g_{a(bcd)} \int\limits_{\nu, \bm k} [C_+\Sigma_-^{(2)}C_+]_{cd}(\nu, \bm k) = 0.
\end{align} 
We can see that this vanishes without performing the integral, as $C_+\Sigma_-^{(2)}C_+ = C_+ - C_+^T$ is anti-symmetric, and it is contracted with the symmetric indices of $g_{a(bcd)}$.
The one-loop correction to the lower-right quadrant is given as 
\begin{align}
    d^{(2)}_{ab}
    & \equiv
    \parbox{12mm}{
    \centering
    \begin{fmfgraph*}(12,10)
        \setval
        \fmfleft{i}
        \fmfright{o}
        \fmftop{t}
        \fmf{plain}{i,c}
        \fmf{plain}{o,c}
        \fmffreeze
        \fmf{plain,left=.4}{c,a,t,b}
        \fmf{wiggly,left=.4}{b,c}
        \fmfforce{.3w, .75h}{a}
        \fmfforce{.7w, .75h}{b}
        \fmfv{d.f=empty,d.shape=triangle,d.size=2mm}{t}
    \end{fmfgraph*} 
    }
    =
    - 6 g_{c(dab)}  \int\limits_{\nu, \bm k} k^2
    [C_+\Sigma_-^{(2)}G_+]_{cd}(\bm k, \omega)
    = 
    8 \alpha_1 \Gamma \delta_{ab} \integral,&
    \integral &\equiv 
    \int \frac{\dd^d k}{(2\pi)^d}
    k^2 \left(\frac{\alpha_0 + \beta_0 k^2}{r + k^2}\right),
\end{align}
where we have defined the integral $\integral$ for book-keeping.
We will not need explicit evaluation of this integral, because it cancels a contribution from $\Delta^{(2)}$.

Thus, there is no contribution to $\smash{\sigma^{(2)}_{R,\tilde a \tilde b}}$, and the diagrams contributing to $\smash{\sigma_{R, a \tilde b}^{(2)}}$ vanish, while there is a non-zero correction to $\smash{\sigma_{R,ab}^{(2)}}$.
This contribution may seem puzzling at first, as it is non-zero in the $\omega\rightarrow 0$ limit, unlike the bare entropy-consumption vertices.
However, as we will see below, this contribution will cancel with a term from $\Delta^{(2)}$, leaving only corrections $\propto \omega$.
As a result, only the $\smash{\sigma_{R,ab}^{(2)}}$ quadrant of $\smash{\sigma_{R,ij}^{(2)}}$ is non-zero, and $\smash{\Sigma_{R,ij}^{(2)}}$ will have the same form \cite{supp}.


\textit{Appendix C: Calculation of $\Delta^{(4)}$---}%
First, we note that all bare propagators now have the form $\E{\psi_j \psi_j e^{-S_2}}_0\equiv \D_{ij}'$, where the prime indicates the presence of the exponential factor.
These can be interpreted as the leading order time-reversed propagators.
We denote these diagrammatically by a slash, and calculate them as in the linear case above, e.g.
$ C_+'
=
\parbox{8mm}{
\centering
\begin{fmfgraph*}(8,3)
    \setval
    \fmfleft{i}
    \fmfright{o}
    \fmf{corr_p}{i,o}
\end{fmfgraph*}
}
= \big[C^{-1}_+ - \sigma^{(2)}_-\big]^{-1}
$.
This can be obtained either by using the diagrammatic expansion or by explicit matrix inversion \cite{supp}.
The lower-right quadrant of $\Delta_{ij}^{(2)}$ is then
\begin{align}
    &\hskip-1.89mm\Delta^{(2)}_{ab}
    =
     \parbox{16mm}{
    \centering
    \begin{fmfgraph*}(16,10)
        \setval
        \fmfleft{i}
        \fmfright{o}
        \fmftop{t}
        \fmf{corr_p_arrow}{i,c}
        \fmf{corr_p}{c,o}
        \fmffreeze
        \fmf{corr_p_end,left}{c,t}
        \fmf{plain,left}{t,c}
        \fmfdot{c}
    \end{fmfgraph*}
    }
    +
    \parbox{16mm}{
    \centering
    \begin{fmfgraph*}(16,10)
        \setval
        \fmfleft{i}
        \fmfright{o}
        \fmftop{t}
        \fmf{corr_p}{i,c}
        \fmf{corr_p_arrow}{o,c}
        \fmffreeze
        \fmf{corr_p_end,left}{c,t}
        \fmf{plain,left}{t,c}
        \fmfdot{c}
    \end{fmfgraph*}
    }
    +
    \parbox{16mm}{
    \centering
    \begin{fmfgraph*}(16,10)
        \setval
        \fmfleft{i}
        \fmfright{o}
        \fmftop{t}
        \fmf{corr_p}{i,c,o}
        \fmffreeze
        \fmf{corr_p_end,left}{c,t}
        \fmf{fermion,left}{t,c}
        \fmfdot{c}
    \end{fmfgraph*}
    }
    + \cdots
    \!\!\overset{\text{1-loop}}{\sim}\!\!
    \parbox{20mm}{
    \begin{fmfgraph*}(20,2)
        \setval
        \fmfleft{i}
        \fmfright{o}
        \fmf{plain}{i,c1}
        \fmf{plain}{c2,o}
        \fmf{plain}{c1,v,c2}
        \fmfsquare{v}
        \fmfshade{c1}
        \fmfshade{c2}
    \end{fmfgraph*}
    }
    +
    \parbox{30mm}{
    \begin{fmfgraph*}(30,2)
        \setval
        \fmfleft{i}
        \fmfright{o}
        \fmf{plain}{i,c1}
        \fmf{plain}{c3,o}
        \fmf{plain}{c1,v1,c2,v2,c3}
        \fmfsquare{v1}
        \fmfsquare{v2}
        \fmfshade{c1}
        \fmfshade{c2}
        \fmfshade{c3}
    \end{fmfgraph*}
    }
    + \cdots.
\end{align}
Here, we have defined the renormalizd primed propagator, 
$
    C_R'
    =
    \parbox{10mm}{
    \begin{fmfgraph*}(10,3.8)
        \setval
        \fmfleft{i}
        \fmfright{o}
        \fmf{plain}{i,c,o}
        \fmfshade{c}
    \end{fmfgraph*}
    },
    $
in relation to $C_R$, in the same way as $C$ is defined in relation to $C'$, only with $\smash{\sigma_R^{(2)}}$ instead of the bare vertex.
We have chosen to use the renormalized $C$ and $\sigma^{(2)}$, which is justified as the differences are of two-loop order (hence only renormalized quantities appear).
When doing this, we find expressions similar to those at tree-level, only with renormalized parameters, in the same way as seen in the case of $C_R$.
We have also defined the new renormalized entropy vertex, as a sum of two-point, one-particle irreducible contributions
\begin{align}
    \sigma_R^{(4)}
    & = 
    \parbox{10mm}{
    \centering
    \begin{fmfgraph*}(10,8)
        \setval
        \fmfleft{i}
        \fmfright{o}
        \fmf{plain}{i,c,o}
        \fmfsquare{c}
    \end{fmfgraph*}
    }
    =
    \parbox{12mm}{
    \centering
    \begin{fmfgraph*}(12,10)
        \setval
        \fmfleft{i}
        \fmfright{o}
        \fmftop{t}
        \fmf{fermion}{i,c}
        \fmf{plain}{c,o}
        \fmffreeze
        \fmf{corr_p_end,left}{c,t}
        \fmf{plain,left}{t,c}
        \fmfdot{c}
    \end{fmfgraph*}
    }
    +
    \parbox{12mm}{
    \centering
    \begin{fmfgraph*}(12,10)
        \setval
        \fmfleft{i}
        \fmfright{o}
        \fmftop{t}
        \fmf{plain}{i,c}
        \fmf{fermion}{c,o}
        \fmffreeze
        \fmf{corr_p_end,left}{c,t}
        \fmf{plain,left}{t,c}
        \fmfdot{c}
    \end{fmfgraph*}
    }
    +
    \parbox{12mm}{
    \centering
    \begin{fmfgraph*}(12,10)
        \setval
        \fmfleft{i}
        \fmfright{o}
        \fmftop{t}
        \fmf{plain}{i,c,o}
        \fmffreeze
        \fmf{corr_p_end,left}{c,t}
        \fmf{fermion,left}{t,c}
        \fmfdot{c}
    \end{fmfgraph*}
    }.
\end{align}
The first two diagrams correspond to the same integral, which we denote as $d^{(3)}$, while the last integral is denoted as $d^{(4)}$.
The diagrams are
\begin{align}
    d^{(3)}_{ab}
    =
    a\,
    \parbox{12mm}{
    \centering
    \begin{fmfgraph*}(12,10)
        \setval
        \fmfleft{i}
        \fmfright{o}
        \fmftop{t}
        \fmf{fermion}{i,c}
        \fmf{plain}{c,o}
        \fmffreeze
        \fmf{corr_p_end,left}{c,t}
        \fmf{plain,left}{t,c}
    \end{fmfgraph*}
    }\, b
    =
    a\,
    \parbox{12mm}{
    \centering
    \begin{fmfgraph*}(12,10)
        \setval
        \fmfleft{i}
        \fmfright{o}
        \fmftop{t}
        \fmf{plain}{i,c}
        \fmf{fermion}{o,c}
        \fmffreeze
        \fmf{corr_p_end,left}{c,t}
        \fmf{plain,left}{t,c}
    \end{fmfgraph*}
    }\, b
    & = 3 \int_{\nu, \bm k} \sigma_{a(bcd)}^{(4)}(\omega) C'_{cd}(\nu)
    = - 4 i\omega \alpha_1 \int_{\bm k} \frac{\epsilon}{r + k^2}
    = - i\omega \delta \alpha_0 \frac{\Gamma}{D} \epsilon.
\end{align}
Here, $\delta \alpha_0$ is the correction to $ \alpha_0 $ from the self energy, and contains the integral $I^{(2)}_d = \int_{\bm k} \frac{ 1 }{ r + k^2 }$; see \cite{supp} for details.
The last one-loop contribution to the deviation from the FDT is
\begin{align}
    d^{(4)}_{ab}
    =
    a\,
    \parbox{12mm}{
    \centering
    \begin{fmfgraph*}(12,10)
        \setval
        \fmfleft{i}
        \fmfright{o}
        \fmftop{t}
        \fmf{plain}{i,c,o}
        \fmffreeze
        \fmf{corr_p_end,left}{c,t}
        \fmf{fermion,left}{t,c}
    \end{fmfgraph*}
    }
    \, b
    &= 6 \int_{\nu, \bm k} \sigma_{c(dab)}^{(4)}(\nu) C'_{cd}(\nu) 
    =
    - 8 \alpha_1 \Gamma \delta_{ab} \integral.
\end{align}
We see that, in fact, $d^{(2)} = - d^{(4)}$ and $d^{(3)} \propto i \omega$, as expected.
We note the appearance of a Dyson series, and define
\begin{align}
    {\Sigma_{R+}^{(4)}}
    & 
    \equiv
    C_{R-}^{-1}  C_{R-}'
    \left[
    {\sigma_{R+}^{(4)}}^{-1}\!
    - {C_{R-}'}
    \right]^{-1}\!
    C_{R-}'C_{R-}^{-1},
\end{align}
such that $\smash{\Delta^{(2)}_{R+,ab} = [C_{R+} {\Sigma_{R-}^{(4)}} C_{R+}]_{ab}}$.
The other quadrants of $\smash{\Delta_{R,ij}^{(2)}}$ follow a similar pattern.
In fact, as $\smash{\sigma^{(4)}_{R,ij}}$ has $\varphi$-external legs exclusively, the lower-right quadrant is the only non-zero component, following the pattern from $\smash{\sigma^{(2)}_{R,ij}}$.
Thus, we define $\smash{\Delta_{R-,ij}^{(2)} = [\D_{R+} \Sigma_{R-}^{(4)} \D_{R+}]_{ij}}$, where $\smash{\Sigma^{(4)}_{R,ij}}$ has the same form as $\smash{\Sigma^{(2)}_{R,ij}}$; it is zero except for the lower-right quadrant given by $\smash[b]{\Sigma^{(4)}_{R, ab}}$.
This can be seen from the diagrammatics, or by performing explicit matrix inversion, as shown in \cite{supp}.
Upon adding the re-summed operators, $ \Sigma_R = \Sigma^{(2)}_R + \Sigma^{(4)}_R$, the contributions from $d^{(2)}$ and $d^{(4)}$ cancel, leaving only that from $d^{(3)}\propto i \omega \delta \alpha_0$.

\end{widetext}
\end{fmffile}

\bibliography{ref,ramin_ref}

\end{document}


\title{Fluctuation Dissipation Relations for the Non-Reciprocal Cahn-Hilliard Model\\{\it Supplemental Material}} 

\author{Martin Kj{\o}llesdal Johnsrud}
\affiliation{Max Planck Institute for Dynamics and Self-Organization (MPI-DS), D-37077 G\"ottingen, Germany}

\author{Ramin Golestanian}
\affiliation{Max Planck Institute for Dynamics and Self-Organization (MPI-DS), D-37077 G\"ottingen, Germany}
\affiliation{Rudolf Peierls Centre for Theoretical Physics, University of Oxford, Oxford OX1 3PU, United Kingdom}

\date{\today}

\maketitle

\tableofcontents

\setcounter{equation}{0}
\setcounter{figure}{0}
\renewcommand{\theequation}{S\arabic{equation}}
\renewcommand{\thefigure}{S\arabic{figure}}

\begin{fmffile}{feyn/supp}

\section{Nonlinear response field theory}
\label{sec:non-linear}

The two-point functions (or ``propagaotrs'') are the correlation function $C = \E{\varphi \varphi}$ and Green's function $G = \E{\varphi i \tilde \varphi}$.
They are the two most important quantities in  Martin-Siggia-Rose Janssen-De Dominicis (MSR-DJ) field theory (also called response field theory).
At linear (or tree) level, the action is in quadratic form.
Using the shorthand notation $\psi_i = (i \tilde \varphi_1, \dots \varphi_1, \dots)$, it is $A_0 = \int \psi_i \D^{-1}_{ij} \psi_j$, where
\begin{align}
    \D^{-1}
    & = 
    \begin{pmatrix}
        -2 q^2 D & G^{-1} \\( G^{-1})^\dagger & 0
    \end{pmatrix} \implies
    \D_{ij} = \E{\psi_i \psi_j}
    =
    \begin{pmatrix}
        0& G^\dagger \\ G & 2 q^2 D G G^\dagger
    \end{pmatrix},
\end{align}
so $C = 2 q^2 D G G^\dagger$.
Notice that this is a consequence of the lower quadrant of $\D^{-1}$ being zero.

In fact, this relationship remains true even at nonlinear level.
In this case, we denote the two-point functions with a subscript $R$, $C_R$ and $G_R$, for ``renormalized''.
These can be calculated perturbatively using diagrammatic methods, where vertices symbolize nonlinearities, and the edges connecting them are the propagators~\cite{tauberCriticalDynamicsField2014,hertzPathIntegralMethods2016b}.
These diagrams yield the vertex functions $\Gamma^{(n,m)}$, as the sum of all amputated, one-particle irreducible graphs, and
\begin{align}
    \D_R^{-1} 
    =
    \begin{pmatrix}
        \Gamma^{(2,0)} & \Gamma^{(1,1)}\\ {\Gamma^{(1,1)}}^\dagger & \Gamma^{(0,2)}
    \end{pmatrix}.
\end{align}
We note that, as $G(t)\propto \theta(t)$ and $\theta(t < 0) = 0$, any closed loop consisting of only $G$ vanishes.
To see this, we note that such a diagram (in real space) takes the form $G(t_1-t_2)G(t_2-t_3)\dots G(t_n-t_1)$, which means at least one of the arguments must at all times be less than zero.
Therefore, closed response loops vanish as a consequence of causality.
$\Gamma^{(0,2)}$  must contain closed response-loops, and therefore vanish.
From these considerations, we obtain the following relations
\begin{align}
    G_R^{-1} & = \left(\Gamma^{(1,1)}\right)^{-1} 
    = G_0^{-1} - \Pi^{(1,1)}, &
    -2 q^2 D_R & = \Gamma^{(2,0)} = - 2 q^2 D_0 - \Pi^{(2,0)}.
\end{align}
Here, $\Pi^{(n,m)}$ are the \textit{self-energies}, i.e. one-loop-irreducible corrections to the propagators.
We thus find
\begin{align}
    C_R^{-1} 
    & = -\left( \Gamma^{(1,1)}\right)^{-1} \Gamma^{(2,0)} \left({\Gamma^{(1,1)}}^\dagger\right)^{-1}
    = 2 q^2 D_R G_R G_R^\dagger.
\end{align}
At one loop, there are no diagrams that contribute to $\Pi^{(2,0)}$ as the four-point vertex only has one response field leg.
This means that $C_R$ is given, to one loop, by the correction to $G_R$ only.
In fact, to all orders in the perturbative expansion, corrections to $D_R$ are of higher order in momentum, such that $D_R = D + \Oh(q^2)$.
This is a consequence of the conservation law.

\section{Bubble diagrams}

Following the Feynman rules, the first bubble diagram is
\begin{align}
    \parbox{10mm}{
    \centering
    \begin{fmfgraph*}(5,5)
        \setval
        \fmfleft{i}
        \fmfright{o}
        \fmf{plain,left}{i,o}
        \fmf{plain,left}{o,i}
        \fmfdot{o}
    \end{fmfgraph*}}
    = \E{-S}
    &=
    \frac{1}{2} 
    \int\limits_{\nu, \bm k} \sigma_{ab}(\nu) \E{\varphi_a(\nu, \bm k) \varphi_b(-\bm k, -\nu) }\\
    & = 
    \frac{1}{2} (2\pi)^{d+1} \delta(\omega = 0)\delta^d(\bm q = 0)\int\limits_{\nu, \bm k}\sigma_{ab}(\nu) C_{ab}(\nu, \bm k) \\
    & = - 2 T L^d \int\limits_{\bm k} \frac{\alpha_0^2 \Gamma k^2}{k^2 + r}
    \sim - 2\alpha_0^2 \Gamma (T L^d \Lambda^d), \quad T, L ,\Lambda\rightarrow \infty.
\end{align}
Here, $L$ is the linear dimension of the system, $T$ is the total time, and $\Lambda$ is the short wavelength cutoff.
The bubble diagram thus has both IR and UV divergences. This is to be expected as the entropy production in steady state should be proportional to the number of degrees of freedom, which diverges with $T$, $L$, and $\Lambda$. However, this divergence is a feature of the perturbative expansion, and can be eliminated via appropriate re-arrangement of the full sum, as follows
\begin{align}\nonumber
    \Delta_{a\tilde b}
    & =
    \parbox{12mm}{
    \begin{fmfgraph*}(12,5)
        \setval
        \fmfleft{i}
        \fmfright{o}
        \fmf{plain}{i,c}
        \fmf{plain,tension=2}{c,k}
        \fmf{wiggly,tension=2}{k,o}
        \fmfdot{c}
    \end{fmfgraph*}
    }
    +
    \parbox{12mm}{
    \centering
    \begin{fmfgraph*}(5,5)
        \setval
        \fmfleft{i}
        \fmfright{o}
        \fmf{plain,left}{i,o}
        \fmf{plain,left}{o,i}
        \fmfdot{o}
    \end{fmfgraph*}
    \begin{fmfgraph*}(12,5)
        \setval
        \fmfleft{i}
        \fmfright{o}
        \fmf{plain}{i,k}
        \fmf{photon}{k,o}
    \end{fmfgraph*}
    }
    + 
    \frac{1}{2!}
    \Bigg(
    \parbox{15mm}{
    \begin{fmfgraph*}(15,5)
        \setval
        \fmfleft{i}
        \fmfright{o}
        \fmf{plain}{i,c1}
        \fmf{plain}{c1,c2}
        \fmf{plain,tension=2}{c2,k}
        \fmf{photon,tension=2}{k,o}
        \fmfdot{c1}
        \fmfdot{c2}
    \end{fmfgraph*}
    }
    +
    \parbox{12mm}{
    \centering
    2
    \begin{fmfgraph*}(4,4)
        \setval
        \fmfleft{i}
        \fmfright{o}
        \fmf{plain,left}{i,o}
        \fmf{plain,left}{o,i}
        \fmfdot{o}
    \end{fmfgraph*}
    \\
    \begin{fmfgraph*}(12,5)
        \setval
        \fmfleft{i}
        \fmfright{o}
        \fmf{plain}{i,c}
        \fmf{plain,tension=2}{c,k}
        \fmf{wiggly,tension=2}{k,o}
        \fmfdot{c}
    \end{fmfgraph*}}
    +
    \parbox{22mm}{
    \centering
    \begin{fmfgraph*}(4,4)
        \setval
        \fmfleft{i}
        \fmfright{o}
        \fmf{plain,left}{i,o}
        \fmf{plain,left}{o,i}
        \fmfdot{o}
        \fmfdot{i}
    \end{fmfgraph*}
    +
    2
    ${
    \begin{fmfgraph*}(4,4)
        \setval
        \fmfleft{i}
        \fmfright{o}
        \fmf{plain,left}{i,o}
        \fmf{plain,left}{o,i}
        \fmfdot{o}
    \end{fmfgraph*}}^2$
    \\
    \begin{fmfgraph*}(14,5)
        \setval
        \fmfleft{i}
        \fmfright{o}
        \fmf{plain}{i,k}
        \fmf{photon}{k,o}
    \end{fmfgraph*}
    }
    \Bigg)\\ \nonumber
    & \quad +
    \frac{1}{3!}
    \Bigg(
    \parbox{20mm}{
    \begin{fmfgraph*}(20,5)
        \setval
        \fmfleft{i}
        \fmfright{o}
        \fmf{plain}{i,c1}
        \fmf{plain}{c1,c2}
        \fmf{plain}{c2,c3}
        \fmf{plain,tension=2}{c3,k}
        \fmf{photon,tension=2}{k,o}
        \fmfdot{c1}
        \fmfdot{c2}
        \fmfdot{c3}
    \end{fmfgraph*}
    }
    +
    3\times
    \parbox{15mm}{
    \centering
    \begin{fmfgraph*}(4,4)
        \setval
        \fmfleft{i}
        \fmfright{o}
        \fmf{plain,left}{i,o}
        \fmf{plain,left}{o,i}
        \fmfdot{o}
    \end{fmfgraph*}
    \begin{fmfgraph*}(15,5)
        \setval
        \fmfleft{i}
        \fmfright{o}
        \fmf{plain}{i,c1}
        \fmf{plain}{c1,c2}
        \fmf{plain,tension=2}{c2,k}
        \fmf{photon,tension=2}{k,o}
        \fmfdot{c1}
        \fmfdot{c2}
    \end{fmfgraph*}
    }
    +
    3\times
    \parbox{22mm}{
    \centering
    \begin{fmfgraph*}(4,4)
        \setval
        \fmfleft{i}
        \fmfright{o}
        \fmf{plain,left}{i,o}
        \fmf{plain,left}{o,i}
        \fmfdot{o}
        \fmfdot{i}
    \end{fmfgraph*}
    +
    2
    ${
    \begin{fmfgraph*}(4,4)
        \setval
        \fmfleft{i}
        \fmfright{o}
        \fmf{plain,left}{i,o}
        \fmf{plain,left}{o,i}
        \fmfdot{o}
    \end{fmfgraph*}}^2$
    \\
    \begin{fmfgraph*}(14,5)
        \setval
        \fmfleft{i}
        \fmfright{o}
        \fmf{plain}{i,c}
        \fmf{plain,tension=2}{c,k}
        \fmf{wiggly,tension=2}{k,o}
        \fmfdot{c}
    \end{fmfgraph*}
    }
    +
    \parbox{36mm}{
    \centering
    {$\Big[
    \begin{fmfgraph*}(4,4)
        \setval
        \fmfleft{i}
        \fmfright{o1,o2}
        \fmf{plain,right=1/2}{i,o1}
        \fmf{plain,right=1/2}{o1,o2}
        \fmf{plain,right=1/2}{o2,i}
        \fmfdot{o1}
        \fmfdot{o2}
        \fmfdot{i}
    \end{fmfgraph*}
    +
    3\,
    \begin{fmfgraph*}(4,4)
        \setval
        \fmfleft{i}
        \fmfright{o}
        \fmf{plain,left}{i,o}
        \fmf{plain,left}{o,i}
        \fmfdot{o}
        \fmfdot{i}
    \end{fmfgraph*}
    \,\,
    \begin{fmfgraph*}(4,4)
        \setval
        \fmfleft{i}
        \fmfright{o}
        \fmf{plain,left}{i,o}
        \fmf{plain,left}{o,i}
        \fmfdot{o}
    \end{fmfgraph*}
    + 3!\,
    {
    \begin{fmfgraph*}(4,4)
        \setval
        \fmfleft{i}
        \fmfright{o}
        \fmf{plain,left}{i,o}
        \fmf{plain,left}{o,i}
        \fmfdot{o}
    \end{fmfgraph*}}^3
    \Big]$}
    \\
    \begin{fmfgraph*}(14,5)
        \setval
        \fmfleft{i}
        \fmfright{o}
        \fmf{plain}{i,k}
        \fmf{photon}{k,o}
    \end{fmfgraph*}
    }
    \Bigg)
    +\dots \\ \nonumber
    & = 
    -\,
    \parbox{12mm}{
    \centering
    \begin{fmfgraph*}(12,5)
        \setval
        \fmfleft{i}
        \fmfright{o}
        \fmf{plain}{i,k}
        \fmf{photon}{k,o}
    \end{fmfgraph*}
    }
    +
    \Bigg(
    \parbox{12mm}{
    \centering
    \begin{fmfgraph*}(12,5)
        \setval
        \fmfleft{i}
        \fmfright{o}
        \fmf{plain}{i,k}
        \fmf{photon}{k,o}
    \end{fmfgraph*}
    } 
    +
    \parbox{12mm}{
    \begin{fmfgraph*}(12,5)
        \setval
        \fmfleft{i}
        \fmfright{o}
        \fmf{plain}{i,c}
        \fmf{plain,tension=2}{c,k}
        \fmf{wiggly,tension=2}{k,o}
        \fmfdot{c}
    \end{fmfgraph*}
    }
    +
    \parbox{15mm}{
    \begin{fmfgraph*}(15,5)
        \setval
        \fmfleft{i}
        \fmfright{o}
        \fmf{plain}{i,c1}
        \fmf{plain}{c1,c2}
        \fmf{plain,tension=2}{c2,k}
        \fmf{photon,tension=2}{k,o}
        \fmfdot{c1}
        \fmfdot{c2}
    \end{fmfgraph*}
    }
    +
    \parbox{20mm}{
    \begin{fmfgraph*}(20,5)
        \setval
        \fmfleft{i}
        \fmfright{o}
        \fmf{plain}{i,c1}
        \fmf{plain}{c1,c2}
        \fmf{plain}{c2,c3}
        \fmf{plain,tension=2}{c3,k}
        \fmf{photon,tension=2}{k,o}
        \fmfdot{c1}
        \fmfdot{c2}
        \fmfdot{c3}
    \end{fmfgraph*}
    }
    + \dots
    \Bigg)
    \\
    & \quad\quad\quad\quad\quad\quad
    \times
    \Bigg(
    \parbox{5mm}{
    \begin{fmfgraph*}(5,5)
        \setval
        \fmfleft{i}
        \fmfright{o}
        \fmf{plain,left}{i,o}
        \fmf{plain,left}{o,i}
        \fmfdot{o}
    \end{fmfgraph*}
    }
    +
    \frac{1}{2!}
    \Big[ 
    \parbox{18mm}{
    \begin{fmfgraph*}(4,4)
        \setval
        \fmfleft{i}
        \fmfright{o}
        \fmf{plain,left}{i,o}
        \fmf{plain,left}{o,i}
        \fmfdot{o}
        \fmfdot{i}
    \end{fmfgraph*}
    +
    2
    ${
    \begin{fmfgraph*}(4,4)
        \setval
        \fmfleft{i}
        \fmfright{o}
        \fmf{plain,left}{i,o}
        \fmf{plain,left}{o,i}
        \fmfdot{o}
    \end{fmfgraph*}}^2$
    }
    \Big]
    +
    \frac{1}{3!}
    \parbox{36mm}{
    \centering
    {$\Big[
    \begin{fmfgraph*}(4,4)
        \setval
        \fmfleft{i}
        \fmfright{o1,o2}
        \fmf{plain,right=1/2}{i,o1}
        \fmf{plain,right=1/2}{o1,o2}
        \fmf{plain,right=1/2}{o2,i}
        \fmfdot{o1}
        \fmfdot{o2}
        \fmfdot{i}
    \end{fmfgraph*}
    +
    3\,
    \begin{fmfgraph*}(4,4)
        \setval
        \fmfleft{i}
        \fmfright{o}
        \fmf{plain,left}{i,o}
        \fmf{plain,left}{o,i}
        \fmfdot{o}
        \fmfdot{i}
    \end{fmfgraph*}
    \,\,
    \begin{fmfgraph*}(4,4)
        \setval
        \fmfleft{i}
        \fmfright{o}
        \fmf{plain,left}{i,o}
        \fmf{plain,left}{o,i}
        \fmfdot{o}
    \end{fmfgraph*}
    + 3!\,
    {
    \begin{fmfgraph*}(4,4)
        \setval
        \fmfleft{i}
        \fmfright{o}
        \fmf{plain,left}{i,o}
        \fmf{plain,left}{o,i}
        \fmfdot{o}
    \end{fmfgraph*}}^3
    \Big]$}}
    +\dots
    \Bigg).
    \label{eq: diag expansion}
\end{align}
The $1/n!$ prefactors come from the expansion of the exponential, while the other numerical factors are combinatorial factors that appear in the construction of the connected parts of the disconnected diagrams.
Upon inspection of the factorized vacuum bubbles, we can verify that they correspond to $\E{e^{-S[\varphi]}} = 1$ exactly.

\section{Detailed calculation of the FDR}

We provide the explicit verification of the bottom-left identity, as given in Eq.~(4) in the main paper, for the linear NRCH field theory. The propagators in Fourier space are given as
\begin{align}
    G_+
    = [(-i\omega+ q^2 \Es ) \one + q^2 \Gamma  \A \epsilon]^{-1}
    =
    \frac{1}{\gamma}\left[(-i\omega + q^2 \Gamma \Es) \one - q^2 \Gamma \A \epsilon\right],
\end{align}
where $\Es(q) = q^2 + r$, $\A(q) = \alpha_0 + \beta_0 q^2$, and $\gamma = (-i\omega + q^2 \Gamma \Es)^2 + q^4\Gamma^2 \A^2$.
Furthermore, 
\begin{align}
    C_+ = 2 q^2 D G_+ G_+^\dagger
    = \frac{ 2 q^2 D }{|\gamma|^2}
    \left\{\left[ \omega^2 + q^4 \Gamma^2 \left(\A^2 + \Es^2\right) \right] \one - 2 i \omega q^2 \Gamma \A \epsilon \right\},
\end{align}
and
\begin{align}
    \frac{1}{2}
    \left( G_+ - G_-  \right)
    = i \I G
    & = - \frac{ i \omega }{|\gamma|^2}
    \left\{
        \left[-\omega^2 +q^4\Gamma^2\left(\A^2 - \Es^2 \right)\right]\one
        - 2 q^4 \Gamma^2 \Es \A \epsilon
    \right\},
\end{align}
so the left-hand side of the FDT, Eq.~(1), is found as
\begin{align}
    \label{eq: FDT tree level}
     G_+ - G_-   - \frac{i \omega}{q^2 D} C_+
    = 
    -\frac{4 i\omega q^2\Gamma \A}{|\gamma|^2}
    \left[q^2\Gamma \A \one + (-i\omega + q^2\Gamma \Es) \epsilon\right].
\end{align}
As expected, the above result vanishes for $\alpha_0 = 0$.
For low $\alpha_0$ and $\beta_0 = 0$, the leading order violation of the FDT is 
\begin{align}
     G_+ - G_-   - \frac{i \omega}{q^2 D} C_+
    \sim
    - 4 i \omega q^2 \Gamma \alpha_0 \frac{-i\omega + q^2\Gamma \Es}{(\omega^2 + q^4 \Gamma^2 \Es^2)^2}\epsilon  + \Oh(\alpha_0^2).
\end{align}
This corresponds to the $\Oh(\alpha_0)$ contribution of the first diagram in Eq.~\eqref{eq: diag expansion}, $C_-\sigma_+G_-$, showing the order-by-order consistency of the expansion.

Calculating $\Sigma_+ = [\sigma_+^{-1} - C_-]^{-1}$ explicitly, we obtain
\begin{align}
    \Sigma 
    &= 
    \frac{8 (\omega q^2  \Gamma \A)^2}{q^2 D|\gamma|^2}
    \left[ (\omega^2 + q^4 \Gamma^2 \A^2 + q^4\Gamma^2 \Es^2)\one - 2i \omega q^2\Gamma \A \epsilon \right]
    + \frac{2 i \omega q^2 \Gamma \A}{D} \epsilon.
\end{align}
Finally, we find
\begin{align}
    \Delta_{+,a \tilde b}
    =
    (C_- \Sigma_+ G_-)_{ab}
    =
    -
    \frac{4 i\omega q^2 \Gamma \A}{|\gamma|^2}
    \left[q^2 \Gamma \A \one + (-i\omega + q^2\Gamma \Es) \epsilon\right]_{ab},
\end{align} 
which is identical to Eq.~\eqref{eq: FDT tree level}, and thus verifies the generalized fluctuation dissipation theorem.
We have found a second relation for $\Delta_{a\tilde b}$ in the case of the linear theory, and indeed an explicit calculation shows that
\begin{align}
    \label{eq: FDT tree level 2}
     \left[ C_+C_-^{-1} - \one\right]G_-
     = 
     -
    \frac{4 i\omega q^2 \Gamma \A}{|\gamma|^2}
    \left[q^2 \Gamma \A \one + (-i\omega + q^2\Gamma \Es) \epsilon\right].
\end{align}
We see that the right-hand of Eq.~\eqref{eq: FDT tree level} and \eqref{eq: FDT tree level 2} are equal.
Thus, we may equate the left hand sides and rearrange to obtain
\begin{align}
    G_+ - C_+ C_-^{-1} G_- = \frac{i \omega}{q^2 D} C_+,
\end{align}
which is exactly Eq.~(6) from the main text.

In the case where we include only $u$, not $\alpha_1$, then at one-loop, only the $r$-parameter is renormalized. 
This is not a generic property of non-equilibrium theories, but rather, a specific feature for NRCH. 
It is related to the well-known fact that the field renormalization factor in $\varphi^4$-theories is unity at one-loop, due to the symmetry of the interaction-vertex and constraints arising from the conservation law. 
Thus, $\Delta_{ij}$ has the same form, and the above calculations are all valid also at one loop, provided the substitution $r \rightarrow r + \delta r$ is made, where $\delta r$ is the one-loop corrections to $r$.

\section{Renormalization of propagators}

We here find the renormalized propagators $G_R$ and $C_R$, as defined in section \ref{sec:non-linear} to one loop.
This calculation follows the same steps as the familiar calculation of Model B, whose details can be found in~\cite{tauberCriticalDynamicsField2014}, but in more generality.
This has been done in the momentum-shell formalism in~\cite{johnsrudRenromalizationTBP}.
There is one diagram that corrects the self energy of $G$ to one loop order, which is
\begin{align}
    \label{eq: self energy}
    \Pi^{(1,1)}_{ab}(\bm q, \omega) = 
    \parbox{24mm}{
    \centering
    \begin{fmfgraph*}(16,10)
        \setval
        \fmfleft{i}
        \fmfright{o}
        \fmftop{t}
        \fmf{wiggly}{i,c}
        \fmf{plain}{o,c}
        \fmffreeze
        \fmf{plain,left}{c,t}
        \fmf{plain,left}{t,c}
        \fmflabel{$a$}{i}
        \fmflabel{$b$}{o}
    \end{fmfgraph*}
    }
    = 
    - 3 q^2\Gamma g_{a(bcd)} \int\limits_{\nu, \bm k} C_{cd}(\nu, \bm k).
\end{align}
Using the explicit forms of $C$ and $G$ from the main text, we can perform the frequency integral.
We have
\begin{align}
    \int_{ \nu, \bm k} 
    C_{ab}( \nu, \bm k)
    & =
    \delta_{ab} \frac{ D }{ \Gamma } I^{(2)}_d, &
    I^{(2)}_d \equiv \int \frac{\dd^d k}{(2\pi)^d} \, \frac{ 1 }{ \Es(\bm k) }.
\end{align}
This integral is divergent for $d\geq2$ and must be regularized, but for our purposes we may just keep it as an undetermined parameter dependent on the microscopic physics.
The self-energy gives corrections to the linear terms, so $r\rightarrow r + \delta r $ and $\alpha_0 \rightarrow \alpha_0 + \delta \alpha_0$.
In terms of the self energy, $- \Pi_{ab}(\bm q, \omega) = q^2\Gamma (\delta r \delta_{ab} + \delta \alpha_0 \epsilon_{ab})$.
Using
\begin{align}
    3 g_{a(bcd)}\delta_{cd} = U_{ab} \delta_{cd}\delta_{cd} + 2 U_{ab} = 4 U_{ab},
\end{align}
where $U_{ab} = u \delta_{ab} + \alpha_1 \epsilon_{ab}$, we get
\begin{align}
    & \delta r = 4 u \frac{ D }{ \Gamma } I^{(2)}_d, &
    & \delta \alpha_0 = 4 \alpha_1 \frac{ D }{ \Gamma } I^{(2)}_d.
\end{align}

The renormalized propagator is thus
\begin{align}
    G_R^{-1} 
    = G_{0}^{-1} - \Pi
    = \left[ -i \omega + q^2 \Gamma \left(r + \delta r + q^2\right) \right]\one + q^2\Gamma \left(\alpha_0 + \delta \alpha_0 + \beta_0 q^2\right) \epsilon
    \equiv
    \left[ -i \omega + q^2 \Gamma \Es_R \right]\one + q^2 \Gamma \A_R \epsilon.
\end{align}
This propagator is thus exactly the same as $G_0$, only with the substitution $\Es \rightarrow \Es_R = \Es + \delta \Es$, $\A \rightarrow \A_R = \A + \delta \A$,
\begin{align}
    G_R
    \equiv
    \parbox{12mm}{
    \centering
    \begin{fmfgraph*}(12,6)
        \setval
        \fmfleft{i}
        \fmfright{o}
        \fmf{plain}{i,v}
        \fmf{wiggly}{v,o}
        \fmfv{d.f=hatched,d.shape=circle,d.size=3.5mm}{v}
    \end{fmfgraph*}
    }
    = \left[ (-i\omega + q^2 \Gamma \Es_R) + q^2 \Gamma \A_R \epsilon \right]^{-1}
    = \frac{1}{\Delta_R} \left[ (-i\omega + q^2 \Gamma \Es_R) - q^2 \Gamma \A_R \epsilon \right],
\end{align}
where $\Delta_R = (-i\omega + q^2 \Gamma \Es_R )^2 + (q^2 \Gamma \A_R)^2$.

Next, the symmetric propagator is likewise given by the renormalized response propagator $G_R$, and the renormalized diffusion $D_R$.
$D_R$ is renormalized by diagrams with two external $\tilde \varphi$ legs, however due to the conservation law these are higher order in $\bm k$ and thus irrelevant.
Thus, to one loop, we can again just perform the substitution $\Es \rightarrow \Es_R$ and $A\rightarrow \A_R$, yielding
\begin{align}
    C_R 
    \equiv
    \parbox{12mm}{
    \centering
    \begin{fmfgraph*}(12,6)
        \setval
        \fmfleft{i}
        \fmfright{o}
        \fmf{plain}{i,v,o}
        \fmfv{d.f=hatched,d.shape=circle,d.size=3.5mm}{v}
    \end{fmfgraph*}
    }
    = 2 q^2 D G_R G_R^\dagger 
    = \frac{2  q^2 D }{|\Delta_R|^2}
    \left\{ \left[ \omega^2 + q^4\Gamma^2(\Es_R^2 + \A_R^2)\right]\one - 2 i \omega q^2 \Gamma \A_R \epsilon\right\}.
\end{align}

\section{Explicit matrix inversion}

In the text, we obtain two-point entropy-consumption vertices which have the form
\begin{align}
    \sigma_{ij}
    &=
    \begin{pmatrix}
        0 & 0 \\ 0 & \sigma_{ab}
    \end{pmatrix}
\end{align}
for $\sigma^{(2)}$, $\sigma^{(2)}_R$ and $\sigma^{(4)}_R$.
We therefore generically denote them as
\begin{align}
    P_{ij} = 
    \begin{pmatrix}
        0 & 0 \\ 0 & P_{ab}
    \end{pmatrix}.
\end{align}
Here, we show using explicit matrix inversion that for all these cases we obtain the same form for the re-summed operator, i.e. the $\Sigma$'s, where only the lower-right quadrant is non-zero.
The propagator has the form
\begin{align}
    \D^{-1}
    =
    \begin{pmatrix}
        -2 q^2 D & G^{-1}\\{G^\dagger}^{-1} & 0
    \end{pmatrix}
    \implies
    \D
        =
    \begin{pmatrix}
        0 & G^\dagger \\G & C
    \end{pmatrix},\quad
    C = 2q^2 D G G^\dagger.
\end{align}
First, denoting the identity matrix by $I$, we find
\begin{align}
    I - \D P
    &=
    \begin{pmatrix}
        I & G^\dagger \\ 0 & I - C P
    \end{pmatrix}
    \implies
    [I - \D P]^{-1}
    =
    \begin{pmatrix}
        I & - G^\dagger [I - C P]^{-1} \\
        0 & [I - C P]^{-1}
    \end{pmatrix},
\end{align}
which yields
\begin{align}
    Q
    & =
    P [I - \D P]^{-1}
    =
    \begin{pmatrix}
        0 & 0 \\ 0 & [P^{-1} - C]^{-1}
    \end{pmatrix}
    \equiv
    \begin{pmatrix}
        0 & 0 \\ 0 & Q
    \end{pmatrix}.
\end{align}
Setting $P = \sigma_{(R)-}^{(2)}$ and $Q = \Sigma^{(2)}_{(R)-}$ gives the result, as reported in the main text.

Next, for  $\sigma^{(4)}$, we also need the reverse propagators
\begin{align}
    \D'_{ij} = \E{\psi_i \psi_j e^{-S}} 
    \implies
    {\D'}^{-1}
    =
    \begin{pmatrix}
        - 2 q^2 D & G^{-1} \\ {G^\dagger}^{-1} & P
    \end{pmatrix},
\end{align}
where in that case $P = \sigma_-$.
In general, propagators and entropy vertices will have opposite signs in the subscript.
This can be understood as a consequence of one being considered a vertex, such that the sign from $e^{-\frac12 \int \varphi \sigma\varphi}$ is included, whereas for the propagators, deriving from $e^{-\frac12 \int \varphi C^{-1}\varphi}$, this is not the case.
Explicit inversion gives
\begin{align}
    \D' = 
    \begin{pmatrix}
        G^\dagger  Q G & [C' C^{-1} G ]^\dagger \\
        C' C^{-1} G & C'
    \end{pmatrix}, \quad
    C' = [C^{-1} - P]^{-1}, \quad
    Q = [P^{-1} - C]^{-1},
\end{align}
which can also be obtained using diagrammatic methods.
We now seek
\begin{align}
    \Sigma_{R+,ij}^{(4)} 
    & =\left\{
    \D_{R-}^{-1} \D_{R-}'
    \left[
    \sigma_{R+}^{(4)}(I - \D_{R-}'\sigma_{R+}^{(4)})^{-1}
    \right]
    \D_{R-}' \D_{R-}^{-1}
    \right\}_{ij}.
\end{align}
First, we make the following observations
\begin{align}
    I - \D'P
    =
    \begin{pmatrix}
        I & - [C' C^{-1}G^{-1}]^\dagger P  \\ 0 & I - C' P
    \end{pmatrix}
    &\implies
    [I - \D'P]^{-1}
    =
    \begin{pmatrix}
        I & [C' C^{-1}G^{-1}]^\dagger P [I - C'P]^{-1}\\
        0 & [I - C'P]^{-1}
    \end{pmatrix},\\
    &\implies
    Q' =
    P [I - \D'P]^{-1}
    =
    \begin{pmatrix}
        0 & 0\\
        0 & [P^{-1} - C']^{-1}
    \end{pmatrix}
    \equiv
    \begin{pmatrix}
        0 & 0\\
        0 & Q'
    \end{pmatrix}.
\end{align}
Then, using $(C' C^{-1} G)^\dagger = G^\dagger C^{-1} C'$ and $C^{-1} = {G^\dagger}^{-1} (2q^2D)^{-1} G^{-1}$, we obtain
\begin{align}
    \D' Q' \D'
    &=
    \begin{pmatrix}
        G^\dagger Q G & G^\dagger C^{-1} C'\\
        C' C^{-1} G & C'
    \end{pmatrix}
    \begin{pmatrix}
        0 & 0 \\ 0 & Q
    \end{pmatrix}
    \begin{pmatrix}
        G^\dagger Q G & G^\dagger C^{-1} C'\\
        C' C^{-1} G & C'
    \end{pmatrix}\\
    & = 
    \begin{pmatrix}
        G^\dagger C^{-1} C' Q' C' C^{-1} G &&
        G^\dagger C^{-1} C' Q' C'  \\
         C' Q' C' C^{-1} G
         &&
         C'Q' C'
    \end{pmatrix},
\end{align}
and
\begin{align}
    Q
    =
    \D^{-1}\D' Q' \D'\D^{-1}
    &=
    \begin{pmatrix}
        - 2 q^2 D & G^{-1} \\ {G^\dagger}^{-1} & 0
    \end{pmatrix}
    \begin{pmatrix}
        G^\dagger C^{-1} C' Q' C' C^{-1} G &&
        G^\dagger C^{-1} C' Q' C'  \\
         C' Q' C' C^{-1} G
         &&
         C' Q' C'
    \end{pmatrix}
    \begin{pmatrix}
        - 2 q^2 D & G^{-1} \\ {G^\dagger}^{-1} & 0
    \end{pmatrix}\\
    &=
    \begin{pmatrix}
        - 2 q^2 D & G^{-1} \\ {G^\dagger}^{-1} & 0
    \end{pmatrix}
    \bigg(
    \begin{matrix}
        G^\dagger C^{-1} C' Q' C' \overbrace{[C^{-1} G(-2q^2D) + {G^\dagger}^{-1} ]}^{=0}
        &
        G^\dagger C^{-1} C' Q' C' C^{-1}\\
        C' Q C'\underbrace{[C^{-1}G(-2q^2D) + {G^\dagger}^{-1}]}_{=0}
        & 
        C' Q ' C' C^{-1}
    \end{matrix}
    \bigg)
    \\
    & = 
    \bigg(
    \begin{matrix}
        0 &  
        \overbrace{[(-2q^2D) G^\dagger C^{-1} + G^{-1}]}^{=0} C' Q' C' C^{-1}
        \\ 
        0& C^{-1} C' Q' C' C^{-1}
        \\[.1cm]&
    \end{matrix}
    \bigg)
    \equiv 
    \begin{pmatrix}
        0 & 0 \\ 0 & Q
    \end{pmatrix}.
\end{align}
This is a very generic result, and only relies on the assumption that the lower-right quadrant of $P$ is the only nonzero part. The propagator is assumed to be Hermitian, and have vanishing lower-right quadrant.
Thus, identifying $P = \sigma^{(4)}_-$ we find $Q = \Sigma^{(4)}_-$, which has the form discussed in the main text.

\end{fmffile}

\bibliography{ref,ramin_ref}